\documentclass[sigconf,natbib=true]{acmart}
\usepackage{stfloats}
\usepackage{multirow}
\usepackage{subfigure} 
\usepackage{makecell}
\usepackage{caption}
\usepackage{subfigure}
\usepackage{arydshln}
\usepackage{booktabs}
\usepackage{multirow}
\usepackage{tablefootnote}
\usepackage{setspace}
\usepackage{hyperref}
\usepackage{amsmath}
\usepackage{ulem}
\usepackage{multirow}
\usepackage{arydshln}
\usepackage{booktabs}
\usepackage{newfloat}
\usepackage{listings}
\usepackage{bm}
\usepackage[utf8]{inputenc} 
\usepackage[T1]{fontenc}    
\usepackage{hyperref}       
\usepackage{url}            
\usepackage{booktabs}       
\usepackage{amsfonts}       
\usepackage{nicefrac}       
\usepackage{microtype}      
\usepackage{xcolor}         
\usepackage{graphicx}
\usepackage{subfigure}
\usepackage{amsmath}
\usepackage{color}
\usepackage{soul} 
\usepackage{xcolor}
\usepackage{subfigure}
\AtBeginDocument{%
  \providecommand\BibTeX{{%
    \normalfont B\kern-0.5em{\scshape i\kern-0.25em b}\kern-0.8em\TeX}}}

\copyrightyear{2024}
\acmYear{2024}
\setcopyright{rightsretained}
\acmConference[SIGIR '24]{Proceedings of the 47th International ACM SIGIR Conference on Research and Development in Information Retrieval}{July 14--18, 2024}{Washington, DC, USA}
\acmBooktitle{Proceedings of the 47th International ACM SIGIR Conference on Research and Development in Information Retrieval (SIGIR '24), July 14--18, 2024, Washington, DC, USA}
\acmDOI{10.1145/3626772.3657750}
\acmISBN{979-8-4007-0431-4/24/07}

\usepackage{etoolbox}
\makeatletter
\patchcmd{\maketitle}{\@copyrightpermission}{
  \begin{minipage}{0.3\columnwidth}
     \href{https://creativecommons.org/licenses/by/4.0/}{\includegraphics[width=0.90\textwidth]{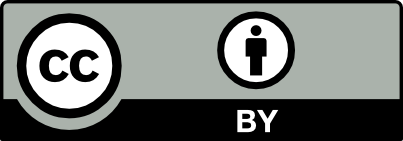}}
  \end{minipage}\hfill
  \begin{minipage}{0.7\columnwidth}
     \href{https://creativecommons.org/licenses/by/4.0/}{This work is licensed under a Creative Commons Attribution International 4.0 License.}
  \end{minipage}
  
  \vspace{5pt}
}{}{}

\makeatother

\begin{document}

\title{Invisible Relevance Bias: Text-Image Retrieval Models Prefer AI-Generated Images}
\author{Shicheng Xu}
\authornote{Equal Contributions}
\affiliation{%
   \institution{CAS Key Laboratory of AI Safety, \\ Institute of Computing Technology, Chinese Academy of Sciences}
  \country{University of Chinese Academy of Sciences, Beijing, China }
}
\email{xschit@163.com}

\author{Danyang Hou}
\authornotemark[1]
\affiliation{%
   \institution{CAS Key Laboratory of AI Safety \\ Institute of Computing Technology, Chinese Academy of Sciences}
  \country{University of Chinese Academy of Sciences, Beijing, China }
}
\email{houdanyang18b@ict.ac.cn}


\author{Liang Pang}
\authornote{Corresponding author}
\affiliation{%
   \institution{CAS Key Laboratory of AI Safety, \\ Institute of Computing Technology, Chinese Academy of Sciences}
  \city{Beijing}
  \country{China}
  }
\email{pangliang@ict.ac.cn}

\author{Jingcheng Deng}
\affiliation{%
   \institution{CAS Key Laboratory of AI Safety, \\ Institute of Computing Technology, Chinese Academy of Sciences}
  \country{University of Chinese Academy of Sciences, Beijing, China }
}
\email{dengjingcheng23s@ict.ac.cn}

\author{Jun Xu}
\affiliation{%
   \institution{Gaoling School of Artificial Intelligence, Renmin University of China}
  \country{Beijing, China}
  }
\email{junxu@ruc.edu.cn}

\author{Huawei Shen}
\affiliation{%
   \institution{CAS Key Laboratory of AI Safety, \\ Institute of Computing Technology,}
  \country{Chinese Academy of Sciences \\ Beijing, China}}
\email{shenhuawei@ict.ac.cn}

\author{Xueqi Cheng}
\affiliation{%
   \institution{\mbox{CAS Key Laboratory of AI Safety, Institute} of Computing Technology, Chinese}
\country{Academy of Sciences, Beijing, China}}
\email{cxq@ict.ac.cn}

\renewcommand{\shortauthors}{Shicheng Xu et al.}

\begin{abstract}

With the application of generation models, internet is increasingly inundated with AI-generated content (AIGC), causing both real and AI-generated content indexed in corpus for search. This paper explores the impact of AI-generated images on text-image search in this scenario. Firstly, we construct a benchmark consisting of both real and AI-generated images for this study. In this benchmark, AI-generated images possess visual semantics sufficiently similar to real images. Experiments on this benchmark reveal that text-image retrieval models tend to rank the AI-generated images higher than the real images, even though the AI-generated images do not exhibit more visually relevant semantics to the queries than real images. We call this bias as \textbf{invisible relevance bias}. This bias is detected across retrieval models with different training data and architectures. Further exploration reveals that mixing AI-generated images into the training data of retrieval models exacerbates the invisible relevance bias. These problems cause a vicious cycle in which AI-generated images have a higher chance of exposing from massive data, which makes them more likely to be mixed into the
training of retrieval models and such training makes the invisible relevance bias more and more serious. To mitigate this bias and elucidate the potential causes of the bias, firstly, we propose an effective method to alleviate this bias. Subsequently, we apply our proposed debiasing method to retroactively identify the causes of this bias, revealing that the AI-generated images induce the image encoder to embed additional information into their representation. This information makes the retriever estimate a higher relevance score. We conduct experiments to support this assertion.

Findings in this paper reveal the potential impact of AI-generated images on retrieval and have implications for further research. Code is released at \url{https://github.com/xsc1234/Invisible-Relevance-Bias}.

\end{abstract}




\maketitle

\section{Introduction} \label{intro}
With the advancement of generation models, the quality of AI-generated content (AIGC) has been increasingly improved~\cite{diffusion,gpt3}. The utilization of AI for content generation has transformed the way of content creation. It not only reduces the cost of content generation but also enhances efficiency, leading to a rapid influx of large amounts of AI-generated content onto the internet~\cite{ai2023information,dai2023llms}.

Information retrieval (IR) is an important way for people to obtain the target information from massive data~\cite{IR}. However, the rapid proliferation of AI-generated content (AIGC) presents a significant new challenge to IR: As the internet becomes increasingly inundated with AI-generated content, the corpus for search contains both real and AI-generated content, so, how will AI-generated content influence the ranking of search? In response to this challenge,~\cite{dai2023llms} conducted a study on text modality and found that neural information retrieval models exhibit a preference for texts generated by Large Language Models. They refer to this category of biases in neural retrieval models towards the LLM-generated text as the \textbf{source bias}. However, beyond textual information, the internet is replete with a substantial number of images, serving as crucial sources for IR systems. Based on this, our paper extends the investigation of source bias in AI-generated content to text-image retrieval models. A pivotal question emerges: What impact will AI-generated images have on existing text-image retrieval models?

\begin{figure}[t]
\centering
	\includegraphics[width=\linewidth]{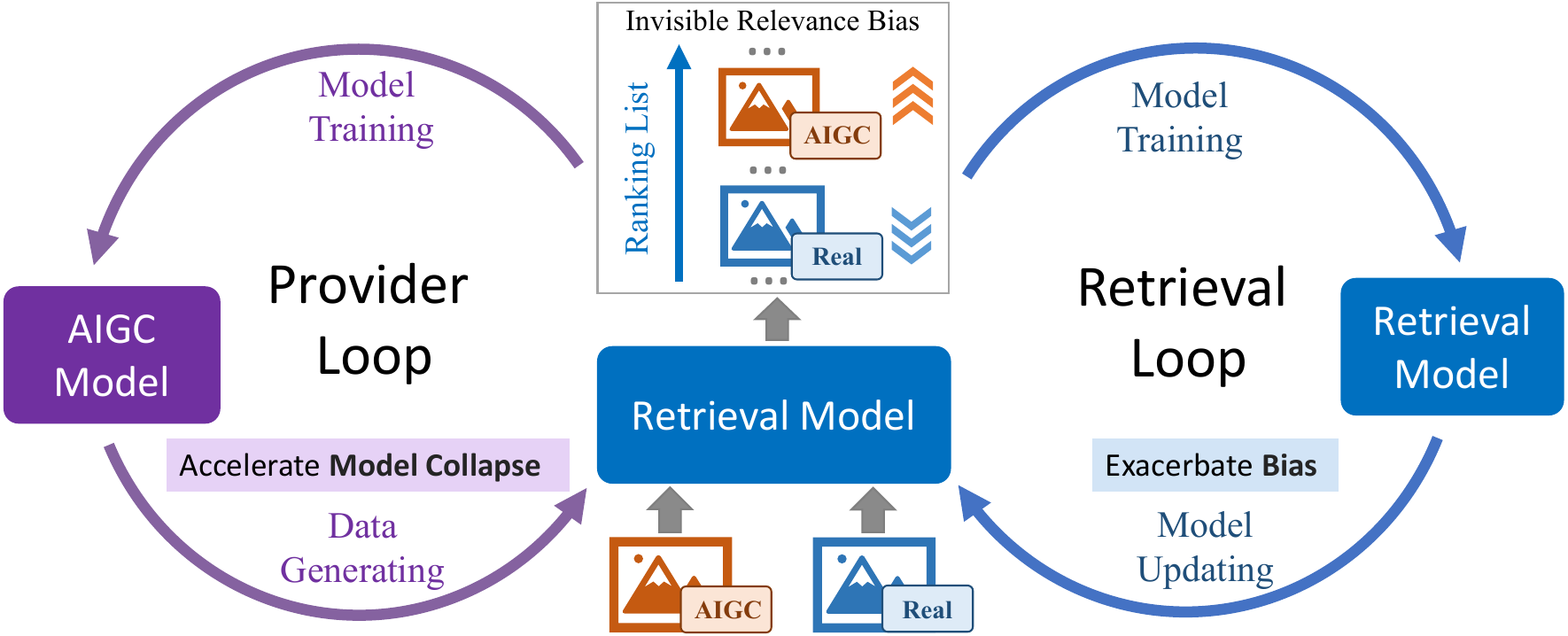}
\caption{Bias found in our paper. IR models tend to rank AI-generated images higher than real images even though they have very similar semantics. This bias increases the likelihood of the generated images being exposed from massive data of internet, which makes them more likely to be mixed into the training of AIGC and retrieval models, leading to more serious bias and forming a vicious cycle.}
        \label{problem}
\end{figure}

A crucial distinction from research on text modality lies in the methodology for \textbf{constructing a benchmark to simulate retrieval scenarios comprising both AI-generated and real images for IR models (§~\ref{benchmark}).}
A reasonable scenario for assessing the potential bias requires that the generated images and the real images have sufficiently similar visual semantics. This can avoid increasing or decreasing some semantic associations between the generated images and the query caused by the image generation. In other words, the IR model preferring (or rejecting) an AI-generated image that is more (or less) semantically relevant to the query than the real image cannot conclusively prove the existence (or nonexistence) of the bias. Study on text modality prompts LLMs to rewrite each real text in the corpus, thereby constructing an LLM-generated text for each real text without introducing additional semantic information. However, this approach is unsuitable for constructing the data in image modality. It is because we find mainstream image generation models such as diffusion models~\cite{rombach2022high} are not good at following prompts to rewrite images while maintaining semantics. Besides, the rewriting paradigm diverges from the prevalent content generation scenario that receives a description as a prompt and generates the texts or images. 
To solve the above problem, we propose an image over-sampling and selection strategy based on the merged caption (§~\ref{Image Generation}). Experimental results and human evaluation show that our proposed method can successfully construct semantically similar AI-generated images for real images (§~\ref{Assessing}). The mixture of these images provides an effective simulation benchmark for investigating text-image retrieval models within scenarios featuring a mix of AI-generated and real images.

Our experiments on the constructed benchmark reveal that \textbf{text-image retrieval models tend to rank the AI-generated images higher than the real images, even though the AI-generated images do not exhibit more visually relevant semantics to the queries than real images (§~\ref{ibb}).} We define this bias as \textbf{invisible relevance bias} introduced by AI-generated images. This bias widely exists in retrieval models with different training data and architectures, including models trained from scratch and models pre-trained on massive image-text pairs, and encompassing dual-encoder and fusion-encoder retrieval models.

Another important point that is not considered in previous work is that AI-generated content does not only have an impact on the inference stage of the retrieval models. Due to the wide distribution of AI-generated images on the internet, they are very likely to be mixed into the training data of retrieval models in the future. Therefore, we further mix the AI-generated images into the training data of the retrieval model and explore the rules of its impact on the retrieval models by adjusting the mixing ratio. Our experiments show that \textbf{as the mixing ratio of AI-generated images in training data increases, the invisible relevance bias becomes more serious (§~\ref{training_bias})}, and the retrieval model exhibits a greater inclination to rank generated images higher. This phenomenon reveals a vicious cycle (Figure~\ref{problem}): Invisible relevance bias causes the generated images to have a higher chance of being obtained from massive
data, which makes them more likely to be mixed into the training of generation and retrieval models and causes the model collapse~\cite{curse}, leading to more serious bias and forming a cycle.

To mitigate the bias, \textbf{we propose an effective training method for debiasing (§~\ref{debias_method})}. Specifically, we introduce a contrastive loss between real and AI-generated images during training. This contrastive loss measures the invisible relevance bias introduced by the AI-generated images for retrieval models. Optimization of this loss can reduce the invisible relevance bias, thereby alleviating the preference to AI-generated images in text-image retrieval models. In addition, we also introduce the sampling probability to enable our debiasing method to dynamically adjust the tolerance to AI-generated images of text-image retrieval models. 

Last but not least, we use our proposed debiasing method to reversely determine that \textbf{the causes of invisible relevance bias is the generated images causing the image encoder to embed additional information into their representation (§~\ref{cause})}. This additional information has a certain consistency in different generated images and can make the retriever estimate a higher relevance score. We also design experiments to support this.

The main contributions of this paper can be concluded as:

(1) We reveal the potential impact of AI-generated images on the ranking results of text-image retrieval systems in the context of the rapid development of AIGC. To reasonably study the impact, we construct a reasonable benchmark to simulate text-image retrieval in scenarios consisting of both real and AI-generated images.

(2) Experimental results show that AI-generated images introduce the \textbf{invisible relevance} to text-image retrieval models, causing the bias that retrieval models prefer ranking AI-generated images higher even though the AI-generated images do not exhibit more visually relevant semantics to the queries than real images. This bias widely exists in retrieval models with different training data and architectures. We also reveal that the loop formed by training and inference causes the retrieval model to fall into a vicious cycle of increasingly serious bias.

(3) We propose an effective method to mitigate the bias by measuring and optimizing the invisible relevance bias introduced by AI-generated images in the training of retrieval models.

(4) We determine the causes of invisible relevance bias is that AI-generated images make the image encoder in the retriever embed additional information to their representations. This additional information is invisible but can amplify the relevant information to get a higher relevance score.



\section{Benchmark Construction} \label{benchmark}
This paper focuses on revealing the potential impact of AI-generated images on the ranking results of text-image retrieval systems. The prerequisite for all this is to construct a retrieval corpus consisting of real images and AI-generated images. This corpus should meet four critical requirements. (\textcolor{red}{\bm{$\mathcal{R}\mbox{-}1$}}) Firstly, from the perspective of fair assessment, a reasonable scenario for assessing the potential bias requires that the generated images and the real images must have sufficiently similar visual semantics. This can avoid increasing or decreasing some semantic associations between the generated images and the queries caused by image generation. That is, the IR model preferring (or rejecting) an AI-generated image that is more (or less) semantically relevant to the query than the real image cannot prove the existence (or nonexistence) of the bias. (\textcolor{red}{\bm{$\mathcal{R}\mbox{-}2$}}) Secondly, retrieval performance on the corpus containing only generated images should not change significantly compared to retrieval performance on real images only. This can further ensure that no additional visual semantics relevant (or irrelevant) to the query are introduced during the image generation. (\textcolor{red}{\bm{$\mathcal{R}\mbox{-}3$}}) Thirdly, the image generation pipeline should be in line with the most common scenario of content generation that receives a description as prompt and generates the texts or images, so that the obtained bias assessment results can be more consistent with the actual scene. (\textcolor{red}{\bm{$\mathcal{R}\mbox{-}4$}}) Fourthly, the number of generated images and real images should be equal to achieve fair comparison. 

\begin{table*}[htbp]
\setlength\tabcolsep{10pt}
\centering
\renewcommand\arraystretch{1.05}
  \caption{Compare generation methods based on the four requirements outlined in Section~\ref{benchmark}. Similarity is the cosine between embeddings of real and generated images encoded by OpenClip. Retrieval Performance is the difference in retrieval performance of BEIT-3 on the corpora only containing generated images and only containing real images respectively.}
  \scalebox{0.75}{
\begin{tabular}{lcccc}
\toprule
                                                        & Similarity (\textcolor{red}{\bm{$\mathcal{R}\mbox{-}1$}}) & Retrieval Performance (\textcolor{red}{\bm{$\mathcal{R}\mbox{-}2$}}) & Generation Pipeline (\textcolor{red}{\bm{$\mathcal{R}\mbox{-}3$}}) & Ratio of Number (\textcolor{red}{\bm{$\mathcal{R}\mbox{-}4$}}) \\ \hline
Single Caption                                          & 0.5275              & $ \left| \triangle \rm{NDCG}@5 \right|=10.8 $      & text to image ($\checkmark$)     & 1:1 ($\checkmark$)  \\ \hdashline
Merged Caption                                          & 0.5348             &  $ \left| \triangle \rm{NDCG}@5 \right|=3.72 $     & text to image ($\checkmark$)     & 1:1 ($\checkmark$)       \\ \hdashline
\makecell[l]{Merged Caption \\ + Image Rewriting}        & 0.5503         & $ \left| \triangle \rm{NDCG}@5 \right|=7.22 $       & text-image to image ($\times$)  & 1:1 ($\checkmark$)    \\ \hdashline
\makecell[l]{Merged Caption \\ + Image Rewriting \\ + Over-Sampling \& Selection} & 0.5845          &  $ \left| \triangle \rm{NDCG}@5 \right|=5.31 $     & text-image to image ($\times$)      & 1:1 ($\checkmark$)   \\ \hdashline
\makecell[l]{Merged Caption \\ + Over-Sampling \& Selection}    & 0.6470       & $ \left| \triangle \rm{NDCG}@5 \right|=1.44 $       & text to image ($\checkmark$)               & 1:1 ($\checkmark$)  \\ \toprule                                 
\end{tabular}
}

  \label{comparision}
\end{table*}

\subsection{Image Generation} \label{Image Generation}

We select two widely used text-image retrieval datasets including Flickr30k~\cite{flickr} and MSCOCO~\cite{coco} as the basis for our benchmark construction. In these two datasets, each image is annotated with five captions that describe the content of the image. For each real image, we aim to generate a corresponding AI-generated image. We propose four image generation methods and use the above four requirements to select the optimal image generation method. Details will be introduced in the following.


\noindent \textbf{\textit{Generation Based on Single Caption.}} In this method, we randomly select one caption from the five captions of each real image and use the selected caption as the prompt of stable diffusion model\footnote{https://huggingface.co/stabilityai/stable-diffusion-xl-base-1.0} to get the AI-generated image corresponding to the real image.

\noindent \textbf{\textit{Generation Based on Merged Caption.}} Since a single caption may not contain the complete visual semantics of the image, we propose to merge five captions to obtain a relatively complete semantic description of the image. Specifically,  for each real image ($I_r$), we use \textit{gpt-3.5-turbo} to combine the five captions to get a newly merged caption $C_m$. And input the merged caption to a stable diffusion model $\mathbf{M}$ to generate the image ($I_g$). The intention of this design is that different captions of an image can be the descriptions from different perspectives of the image. Therefore, merging these captions can obtain an overall description of the image, thereby making stable diffusion generate images that have sufficiently similar visual semantics to the real image. 

\noindent \textbf{\textit{Generation Based on Image Rewriting.}} In addition to the merged caption $C_m$, the real image $I_r$ is also used as the initial image to condition the generation of the new image. Generation constrained by the input real image can output the generated image that has higher similarity to the real image. But it also requires the generation model to have higher multi-modal understanding ability.

\noindent \textbf{\textit{Generation Based on Image Over Sampling and Selection.}} To further narrow the semantic similarity between the generated images and real images, we propose an image over-sampling strategy. Specifically, for a real image $I^r$ we use stable diffusion to perform multiple times generation with different random seeds and get $n$ generated images $\{{I^{g}_1},{I^{g}_{2}},...,{I^{g}_n}\}$. Then, we use the vision encoder $v(\cdot;\theta)$ of a powerful open source pre-trained vision-language model\footnote{https://huggingface.co/laion/CLIP-ViT-H-14-laion2B-s32B-b79K} to get the embedding $\boldsymbol{e_r}$ for $I^r$ and set of embeddings $\boldsymbol{E}=\{\boldsymbol{{e^{g}_1}},\boldsymbol{{e^{g}_2}},...,\boldsymbol{{e^{g}_n}}\}$ for $\{{I^{g}_1},{I^{g}_2},...,{I^{g}_n}\}$. We calculate the cosine similarity between each embedding in $\boldsymbol{E}$ and $\boldsymbol{e_r}$ to get the similarity between the generated images and the real image $I^r$. Finally, we select the generated image with the maximum similarity between $I^r$ as the final generated image ${I}^g$ for the real image.

\noindent \textbf{\textit{Comparison between Different Generation Methods.}} The comparison between different generation methods under the above-mentioned four requirements is shown in Table~\ref{comparision}. Taking these results into account, we choose generation based on merged caption and image over-sampling and selection as the image generation method to construct our benchmark. Using this method, we can get an AI-generated image for each real image in Flicker30k and MSCOCO. In the following content, Flicker30k and MSCOCO indicate the datasets that only contain real images, Flicker30k (AI) and MSCOCO (AI) indicate the datasets that only contain AI-generated images, Flicker30k+AI and MSCOCO+AI indicate the datasets that consist of both real and AI-generated images.


\subsection{Assessing the Quality of Benchmark} \label{Assessing}

Ensuring sufficiently similar visual semantics between generated and real images to avoid increasing or decreasing relevance to the query is a prerequisite for assessing the bias. In this section, we propose two methods to evaluate this. Firstly, we perform retrieval on the corpora that only contain real images and only contain generated images and count the retrieval performance respectively. The intention for this is that if the generated images have more semantics relevant to their corresponding queries than the real images, distinguishing images becomes easier, resulting in significantly higher retrieval performance on the generated images corpus than on the real images corpus. Secondly, we introduce human evaluation to determine whether the generated images have more visual semantics relevant to the queries.

\subsubsection{\textbf{Retrieval Performance}} \label{retrieval_performance}
\begin{table}[t]
  \caption{Retrieval performance (NDCG) on the corpora that only contain real images and only contain AI-generated images. Performance on AI-generated images is not significantly better than the performance on real images can demonstrate the success of our benchmark construction.}
\setlength\tabcolsep{7pt}
\renewcommand\arraystretch{1.0}
\centering
\scalebox{0.75}{
\begin{tabular}{llcccccc}
\toprule
                        &              & \multicolumn{3}{c}{Flicker30k or Flicker30k (AI)}                      & \multicolumn{3}{c}{MSCOCO or MSCOCO (AI)}                 \\ 
                        &              & N@1 & N@3 & N@5 & N@1 & N@3 & N@5 \\ \hline
\multirow{2}{*}{FLAVA}  & Real         & 38.70  & 45.72 & 47.26  & 36.65       &  44.45      & 46.27     \\
                        & AI           & 41.59 & 47.45 & 48.72  & 36.49       &  43.29      & 45.17     \\ \hdashline
\multirow{2}{*}{ALIGIN} & Real         & 45.43 & 50.62 & 51.65  & 38.13  & 44.71    &  46.33       \\
                        & AI           & 43.47 & 49.38 & 50.42  & 36.34    & 43.32       & 45.06          \\ \hdashline
\multirow{2}{*}{BEIT-3} & Real         & 47.45 & 52.15 & 52.87  &  41.24      & 47.16       & 48.63        \\
                        & AI           & 45.31 & 50.49 & 51.43  & 38.33     &  44.76      & 46.19      \\ \toprule
\end{tabular}
}
\label{performance}
\end{table}

The experimental results in Table~\ref{performance} show that retrieval performance on the corpus containing only AI-generated images is not significantly greater than retrieval performance on the corpus containing only real images. It indicates that the AI-generated images in our benchmark do not introduce more visual semantics relevant to the queries. Specifically, we use three open-source and powerful vision-language pre-trained models including FLAVA~\cite{flava}, ALIGN~\cite{align}, and BEIT-3~\cite{beit} to perform retrieval on the corpora that only contain real images and only contain generated images respectively. Since text-image retrieval datasets are composed of real images, the training on these images will introduce additional bias. Therefore, considering the excellent zero-shot text-image retrieval performance of the three models~\cite{beit}, we use these models directly for retrieval in zero-shot setting.

\subsubsection{\textbf{Human Evaluation}} \label{human_evaluation}
Table~\ref{human} shows humans think that in most samples of our benchmark, real images have more or equal visual semantics relevant to the queries than AI-generated images. This further guarantees the fairness of our assessment of invisible relevance bias. If the AI-generated images do not have more relevant visual semantics than real images, while the text-image retrieval model still tends to rank AI-generated images higher than the real images, the invisible relevance bias does exist. Specifically, we invite five humans with master's degrees to participate in the evaluation. We randomly shuffle the dataset into five parts and assign each part randomly to the volunteers. Volunteers are shown with a triple consisting of a caption (i.e., the query), a real image, and its corresponding AI-generated image. We ask them to select which image (real or AI-generated) has more relevant semantics to the caption. We instruct volunteers how to visually judge the relevance between images and captions, and calculate the pass rate through cross-validation between two different volunteers. The formal human evaluation begins when each volunteer's pass rate reaches 95\%. We count the proportion of selections made by humans on our benchmark and the results are shown in Table~\ref{human}.

\begin{table}[t]
  \caption{Proportion of the selections made by humans.}
\setlength\tabcolsep{12pt}
\renewcommand\arraystretch{1.05}
\scalebox{0.75}{
\begin{tabular}{llllll}
\toprule
\multicolumn{3}{c}{Flicker30k+AI} & \multicolumn{3}{c}{MSCOCO+AI} \\ \hline
\multicolumn{6}{c}{Which image is more relevant to the query?}  \\
Real        & AI         & Equal    & Real      & AI        & Equal   \\
46.25\%     & 13.75\%    & 40\%     & 45.35\%   & 12.15\%   & 42.5\% \\ \toprule
\end{tabular}}
\label{human}
\end{table}


\section{Bias Assessment}\label{bias_ass}

Based on the benchmark constructed above, we can assess the impact of AI-generated images on text-image retrieval. We not only study the impact when AI-generated images appear in the corpus, but also further discuss the impact when AI-generated images are mixed into the training data of retrieval models.

\subsection{Text-Image Retrieval Models}

As for training data, our assessment includes not only the retrieval models trained from scratch on supervised text-image pairs but also vision-language models that have been pre-trained on massive image-text pairs. As for model architecture, our assessment includes both dual-encoder-based and fusion-based models. The specific models used in the assessment include: (1) \textbf{NAAF}~\cite{naaf} is a fusion encoder text-image retrieval model that exploits both the positive effect of matched fragments and the negative effect of mismatched fragments to jointly infer text-image similarity. (2) \textbf{VSE}~\cite{vse} is a dual-encoder text-image retrieval model that learns to automatically adapt the best pooling strategy for visual semantic embedding. (3) \textbf{VILT} is a fusion encoder text-image matching model based on the interaction between image and text in Transformer~\cite{tranformer}. It is a vision-language model that has been pre-trained on massive text-image pairs. (4) \textbf{FLAVA} learns strong representations from multimodal. It is a vision-language model pre-trained on massive text-image pairs and can be used as the dual encoder model for text-image retrieval. (5) \textbf{ALIGIN} is also a dual encoder vision-language that has been pre-trained on over one billion image-text pairs. (6) \textbf{BEIT-3} is a multimodal foundation model that has been pre-trained on hundreds of millions of text-image pairs and massive texts and images. It can be used as a dual-encoder text-image retrieval model.

\begin{table*}[htbp]
  \caption{Performance of the retrieval models on the benchmark we constructed consisting of both real and AI-generated images. $\rm{Relative \triangle} > 0$ means retrieval models rank real images higher than AI-generated images, \textcolor{red}{$\rm{Relative \triangle} < 0$} means retrieval models rank AI-generated images higher than real images. The absolute value of $\rm{Relative \triangle}$ indicates the degree of the bias.}
  \label{bias}
\setlength\tabcolsep{6.5pt}
\renewcommand\arraystretch{1.0}
\scalebox{0.7}{
\begin{tabular}{lllccclllccclll}
\toprule
                                &                         &                           & \multicolumn{6}{c}{Flicker30k+AI}                      & \multicolumn{6}{c}{MSCOCO+AI}                 \\ 
                                &                         &                           & NDCG@1 & NDCG@3 & NDCG@5 & R@1    & R@3    & R@5    & NDCG@1 & NDCG@3 & NDCG@5 & R@1 & R@3 & R@5 \\ \hline
\multicolumn{15}{c}{Models trained from scratch}                                                                                                                                         \\ \hdashline
\multirow{3}{*}{Dual-encoder}   & \multirow{3}{*}{VSE}    & Real                      & 16.18       & 26.93       & 29.26       & 26.40       &  56.10      & 65.32       & 11.85       & 20.19       & 22.87       &  19.34   & 42.66    & 53.24    \\
                                &                         & AI-generated              &  19.59      & 29.68       & 31.86       &  31.96      &  59.78      &  68.34      &  13.56      & 20.93      & 23.37       & 22.12    & 43.21    & 53.90     \\
                                &                         & $\rm{Relative} \triangle$ & \textcolor{red}{-17.81}       & \textcolor{red}{-9.00}       & \textcolor{red}{-8.05}       & \textcolor{red}{-17.81}       & \textcolor{red}{-5.8}       & \textcolor{red}{-4.36}     & \textcolor{red}{-13.53}    & \textcolor{red}{-3.64}    & \textcolor{red}{-2.22}  & \textcolor{red}{-13.53}       & \textcolor{red}{-1.29}       & \textcolor{red}{-1.24}        \\ \hdashline
\multirow{3}{*}{Fusion-encoder} & \multirow{3}{*}{NAAF}   & Real                      & 13.40       & 23.39       & 26.14       & 21.86       & 49.41       & 60.28       &  10.61      & 17.73       &  20.45      & 17.30    & 37.26    & 48.02    \\
                                &                         & AI-generated              &  17.04      & 26.04       & 28.31       & 27.79       &  52.70      & 61.70       & 10.75       & 17.87       &  20.33      & 17.54    & 37.50    & 47.24    \\
                                &                         & $\rm{Relative} \triangle$ &  \textcolor{red}{-23.57}      & \textcolor{red}{-10.63}       & \textcolor{red}{-7.86}       & \textcolor{red}{-23.57}       & \textcolor{red}{-6.45}       & \textcolor{red}{-2.31}       & \textcolor{red}{-1.13}       & \textcolor{red}{-0.73}       & 0.62       & \textcolor{red}{-1.13}    & \textcolor{red}{-0.66}     & 1.63    \\ \hline
\multicolumn{15}{c}{Pre-trained Vision-Language Models}                                                                                                                                  \\ \hdashline
\multirow{9}{*}{Dual-encoder}   & \multirow{3}{*}{FLAVA}  & Real                      & 5.44  & 18.44 & 21.79 & 8.88 & 44.92 & 58.14 & 12.59     & 25.98       & 29.02       &  20.54   &  57.30   & 69.34     \\
                                &                         & AI-generated              & 37.61 & 44.86 & 46.36 & 61.33 & 81.34  & 87.26  &  27.01      & 36.81       & 38.87       & 44.06    & 70.99    & 79.12    \\ 
                                &                         & $\rm{Relative} \triangle$ & \textcolor{red}{-148.85}       & \textcolor{red}{-83.78}       & \textcolor{red}{-72.44}      & \textcolor{red}{-148.85}       & \textcolor{red}{-58.32}       & \textcolor{red}{-40.69}       &  \textcolor{red}{-72.81}      & \textcolor{red}{-34.49}       & \textcolor{red}{-29.00}       & \textcolor{red}{-72.81}     & \textcolor{red}{-21.36}    &  \textcolor{red}{-13.21}   \\ \cdashline{2-15} 
                                & \multirow{3}{*}{ALIGIN} & Real                      & 21.92 & 37.20 & 39.05 & 35.76  & 7696 & 84.22 &  18.82      & 31.42        & 33.89       & 30.70    & 64.98     & 74.76    \\
                                &                         & AI-generated              & 25.48 & 39.10 & 40.91 & 41.56  & 78.38 & 85.44 & 21.31       & 33.23       & 35.49       & 34.76     & 67.24    & 76.16    \\
                                &                         & $\rm{Relative} \triangle$ & \textcolor{red}{-14.6}       &  \textcolor{red}{-4.95}      & \textcolor{red}{-4.59}       &  \textcolor{red}{-14.6}      & \textcolor{red}{-1.93}       & \textcolor{red}{-1.49}       & \textcolor{red}{-12.41}      & \textcolor{red}{-5.65}       & \textcolor{red}{-4.63}       & \textcolor{red}{-12.41}    & \textcolor{red}{-3.48}    & \textcolor{red}{-1.88}    \\  \cdashline{2-15} 
                                & \multirow{3}{*}{BEIT-3} & Real                      & 24.37 & 38.67 & 40.50 & 39.76  & 78.22 & 85.46  & 21.38       & 33.26       & 35.57       & 34.88    & 67.11     & 76.22    \\
                                &                         & AI-generated              & 24.40 & 39.54 & 41.12 & 39.80  & 80.50 & 86.68 & 21.24       & 34.55       & 36.63        & 34.64    & 70.86    & 79.08    \\
                                &                         & $\rm{Relative} \triangle$ & \textcolor{red}{-0.72}       & \textcolor{red}{-2.17}       & \textcolor{red}{-1.41}       &  \textcolor{red}{-0.72}      & \textcolor{red}{-2.97}       &  \textcolor{red}{-1.44}      & 0.62       & \textcolor{red}{-3.90}        & \textcolor{red}{-3.01}       &  0.62   & \textcolor{red}{-5.50}    & \textcolor{red}{-3.72}    \\ \hdashline
\multirow{3}{*}{Fusion-encoder} & \multirow{3}{*}{VILT}    & Real                      &  17.53      & 29.63       & 32.16       & 28.60        & 61.90       & 71.90       &  16.30      & 29.71       & 32.08       & 26.60     & 63.10     & 72.50    \\
                                &                         & AI-generated              &  20.04      & 30.43       & 32.71       &  32.70      & 61.30       & 70.30       &  18.29      &  31.21      & 33.50       & 29.85    & 63.30    & 72.30    \\
                                &                         & $\rm{Relative} \triangle$ &  \textcolor{red}{-13.38}      & \textcolor{red}{-2.69}       & \textcolor{red}{-1.69}        & \textcolor{red}{-13.38}       & 0.97       & 2.25       & \textcolor{red}{-11.51}       & \textcolor{red}{-4.90}       &  \textcolor{red}{-4.32}       & \textcolor{red}{-11.51}    &  \textcolor{red}{-0.32}     & 0.28   \\ \toprule
\end{tabular}
}
\end{table*}

\subsection{Experimental Settings and Metrics}
As the neural networks tend to fit the data domain in training~\cite{match-prompt}, our assessment is performed under the out-of-domain setting to try to mitigate potential bias introduced by the domain of the training data. Specifically, for the models that need to train from scratch on supervised text-image pairs (NAAF and VSE), we train them on Flicker30k (MSCOCO) and evaluate their performance on MSCOCO (Flicker30k). For the models that have been pre-trained on massive real text-image pairs and show excellent zero-shot performance in text-image retrieval, we directly use these pre-trained models to perform retrieval on the test datasets. An exception is that even though VILT has been pre-trained, it needs to be combined with a specific multi-layer perceptron to complete the text-image matching task in text-image retrieval. So we fine-tune VILT on supervised text-image retrieval datasets just like NAAF and VSE.

The metric follows~\cite{dai2023llms} to measure the difference between the ranking of real and AI-generated images in the retrieved results as:
\begin{equation}
\rm{Relative} \triangle = \frac{2(\rm{Metric}_{real}-Metric_{AI-generated})}{Metric_{real}+Metric_{AI-generated}} \times 100\%,
\end{equation}
in which Metric can be the metrics for IR such as NDCG@k and R@k. $\rm{Relative \triangle} > 0$ means retrieval models rank real images higher than AI-generated images, $\rm{Relative \triangle} < 0$ means retrieval models rank AI-generated images higher than real images. The absolute value of $\rm{Relative \triangle}$ indicates the degree of the bias~\cite{dai2023llms}.

\subsection{Invisible Relevance Bias}
\label{ibb}

The experimental results are shown in Table~\ref{bias}. Overall, invisible relevance bias widely exists in text-image retrieval models, that is, text-image retrieval models tend to rank AI-generated images higher than real images even though they have very similar visual semantics. Specifically, the following conclusions can be made: (1) The invisible relevance bias exists in both the models trained from scratch and the vision-language models that have been pre-trained on massive supervised text-image pairs. (2) The invisible relevance bias exists in both dual-encoder-based and fusion-encoder-based retrieval models. (3) The invisible relevance bias has a relatively greater impact on the Top-1 retrieved image. In the retrieved list, the Top-1 item is most likely to be clicked by users, which means that invisible relevance bias introduced by AI-generated images will have a huge impact on users’ actual search and click results.

\subsection{More Serious Bias Caused by Training}\label{training_bias}
Due to the wide distribution of AI-generated images on the internet and the bias in Section~\ref{ibb}, AI-generated images are very likely to be mixed into the training data of retrieval models. This section delves deeper into the impact on retrieval performance and invisible relevance bias when AI-generated images are mixed into the training of retrieval models. The experimental results reveal a vicious cycle of falling into more serious invisible relevance bias. Specifically, the invisible relevance bias of text-image retrieval models causes the AI-generated images to have a higher chance of being obtained from massive data, which makes them more likely to be mixed into the training of retrieval models, leading to more serious bias and forming a vicious cycle. This ultimately results in users' search results being surrounded by AI-generated images.

Specifically, we explore the impact of the training mixed with generated images on retrieval by incorporating varying ratios of generated images into the training data. To ensure an accurate assessment, our experiments focus on the model trained from scratch (VSE). This is because pre-trained vision-language models have been pre-trained on massive real text-image pairs, which introduce additional biases in evaluating the impact of specific ratios of generated images in the training data. We reconstruct the training set of Flicker30k by replacing a certain ratio of real images with AI-generated images. For each real image ($I_r$) in the training set, we use the method in Section~\ref{benchmark} to generate its corresponding AI-generated image ($I_g$). Then, we introduce the ratio $\alpha$, which means that in our reconstructed training data, the paired images for $\alpha$ percentage of captions are AI-generated images, and for $(100-\alpha)$ percentage of captions are real images. We change the ratio while keeping the total number of training samples unchanged. We evaluate the performance of the trained model on the test set of Flicker30k+AI (in-domain setting) and MSCOCO+AI (out-of-domain setting). 

\begin{figure*}[t]
    \centering
    \subfigure[$\rm{Relative} \triangle$ on R@k of Flicker30k]{
	\includegraphics[width=1.0in,height=0.8in]{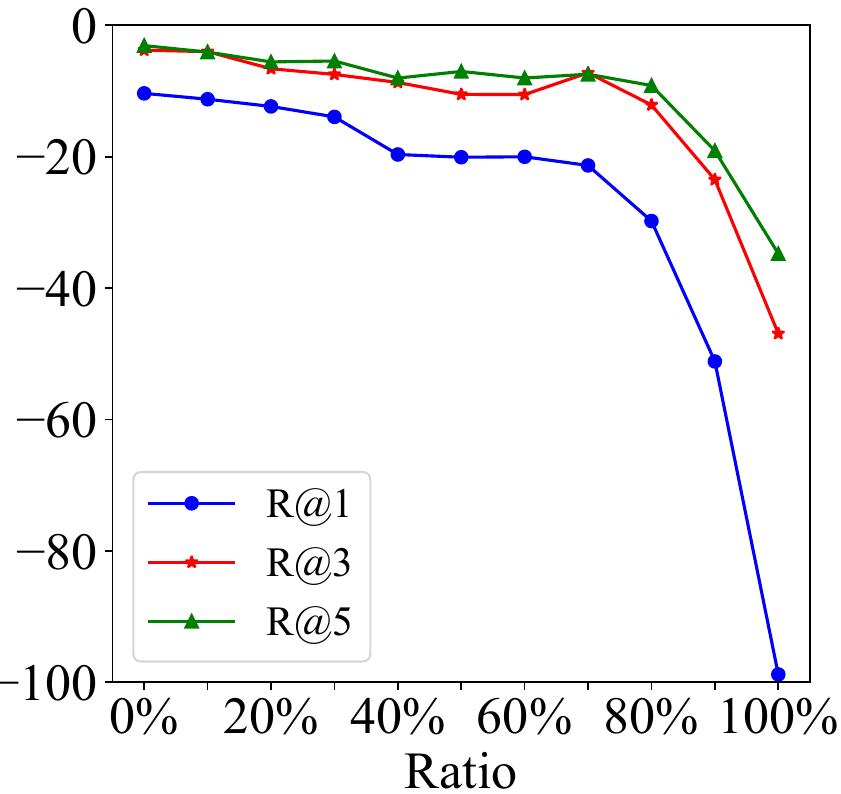}
	\label{ratio_r}
    }\quad
    \subfigure[$\rm{Relative} \triangle$ on NDCG of Flicker30k]{
        \includegraphics[width=1in,height=0.8in]{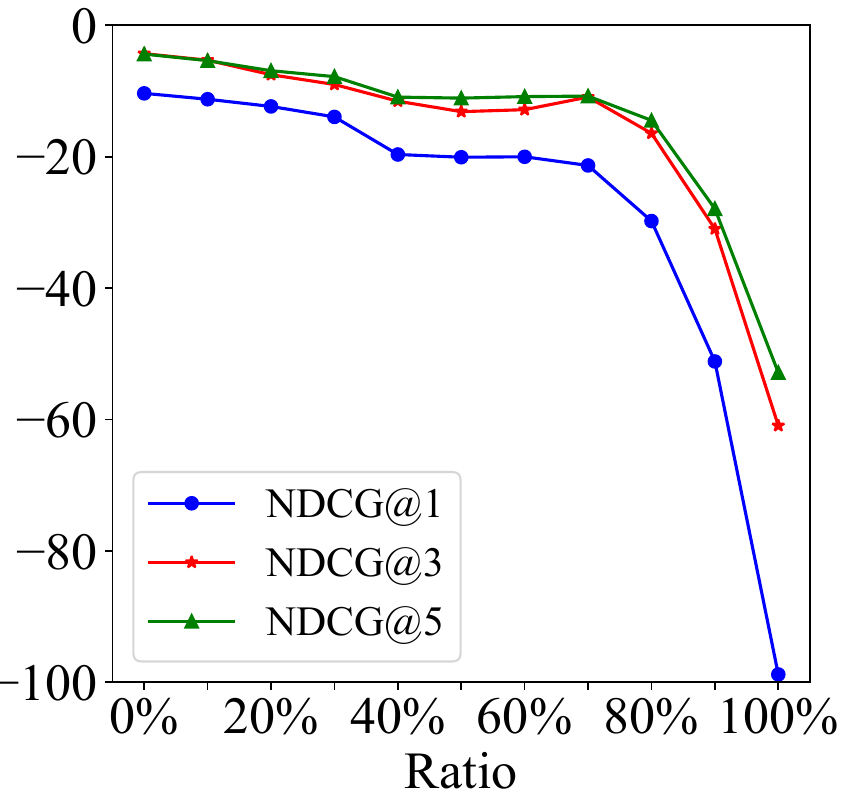}
    \label{ratio_n}
    }\quad
    \subfigure[NDCG on only real or AIGC images of Flicker30k]{
        \includegraphics[width=0.97in,height=0.8in]{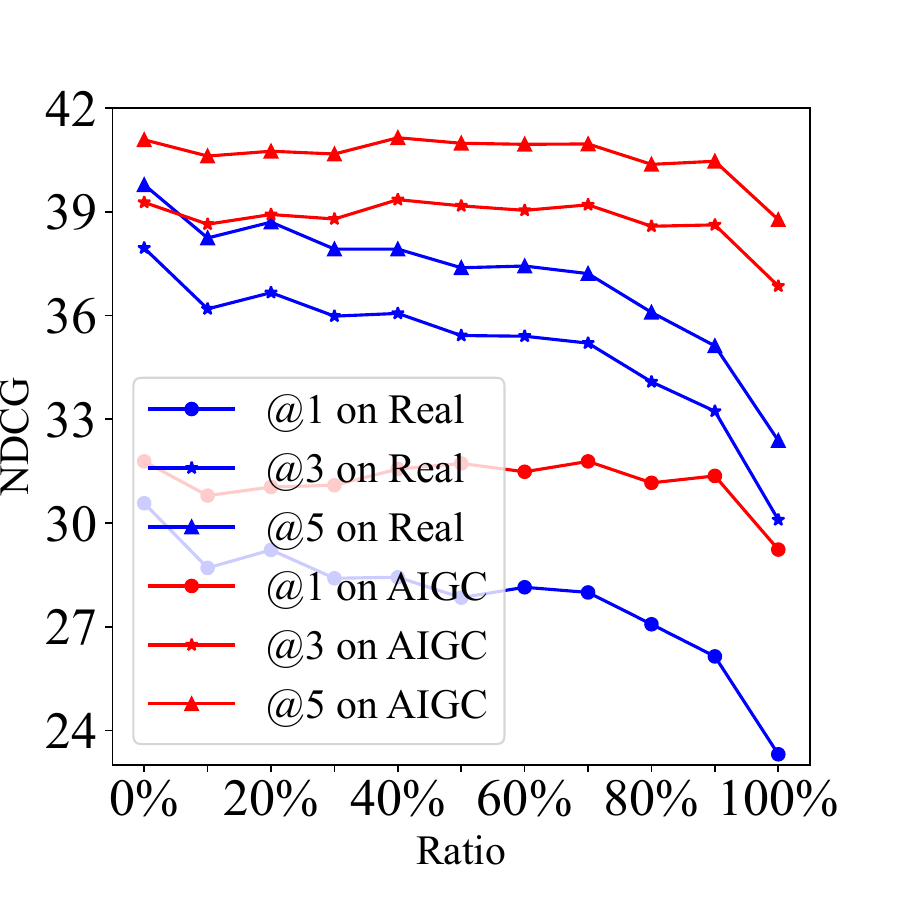}
    \label{ratio_per}
    }\quad
    \subfigure[$\rm{Relative} \triangle$ on R@k of MSCOCO]{
	\includegraphics[width=1in,height=0.8in]{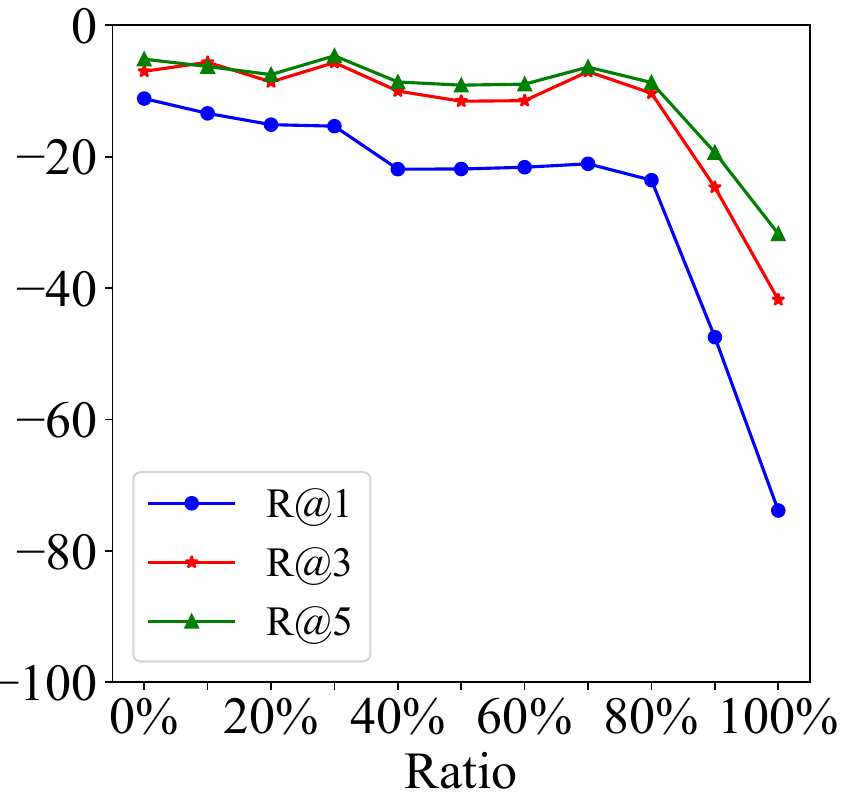}
	\label{ratio_r}
    }\quad
    \subfigure[$\rm{Relative} \triangle$ on NDCG@k of MSCOCO]{
        \includegraphics[width=1in,height=0.8in]{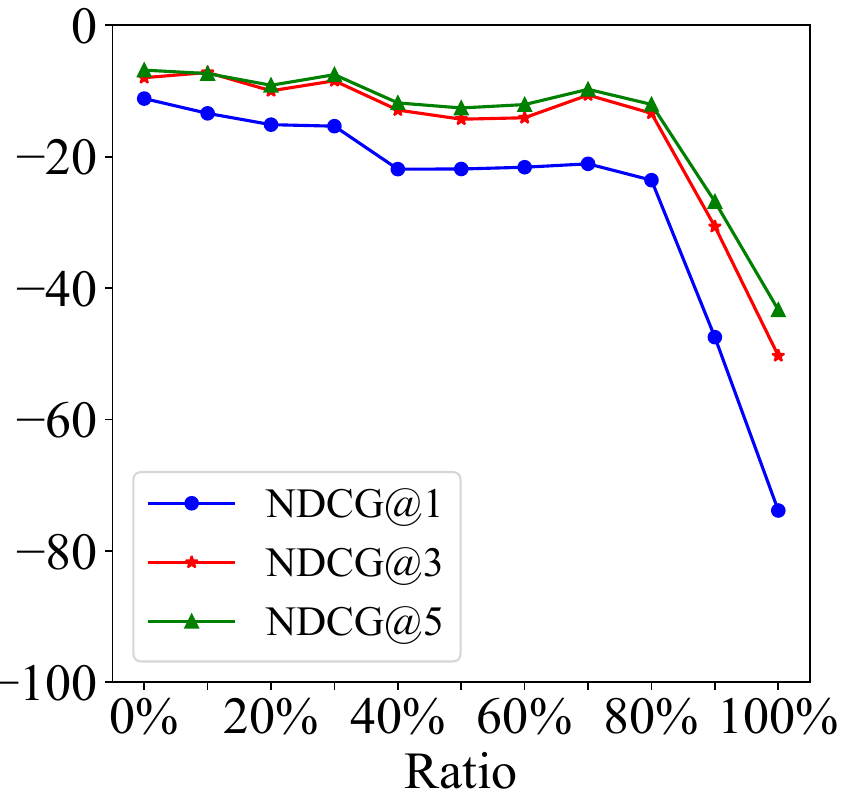}
    \label{ratio_n}
    }\quad
    \subfigure[NDCG on only real or AIGC images of MSCOCO]{
        \includegraphics[width=0.97in,height=0.8in]{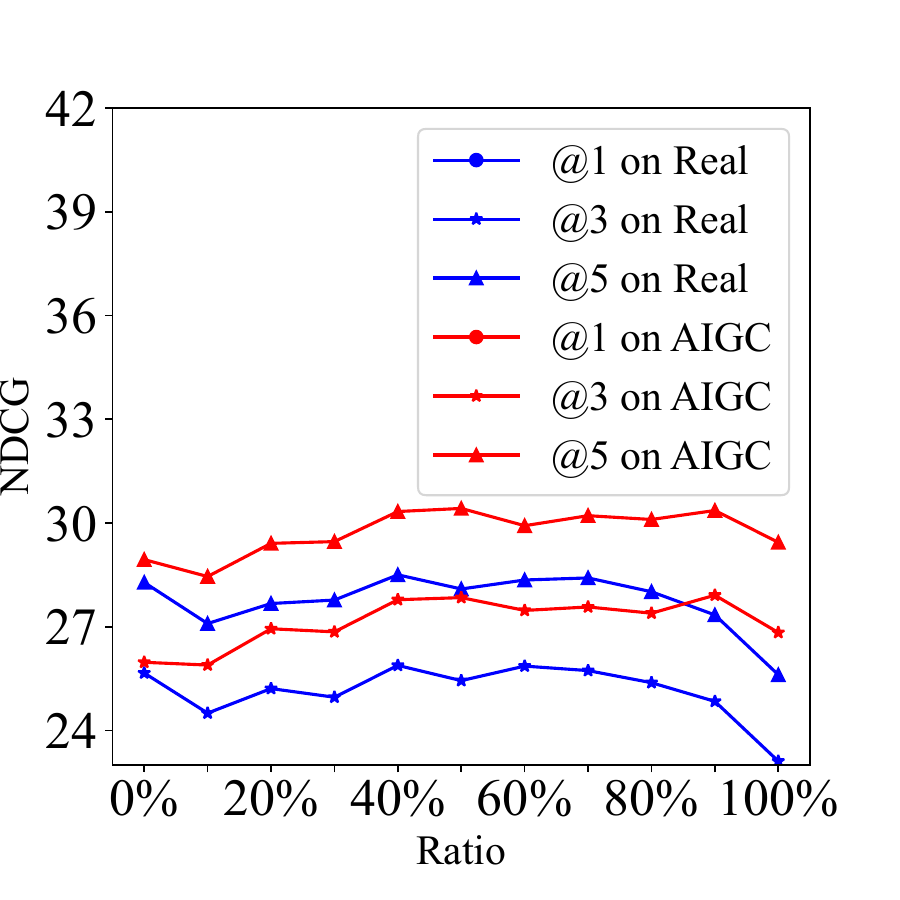}
    \label{ratio_per}
    }
\caption{Assessment results on the training set mixed with AI-generated images. We change the ratio of AI-generated images in the datasets (X-axis) while keeping the total number of training samples unchanged. The model is tested on the test set of Flicker30k+AI (in-domain) and MSCOCO+AI (out-of-domain) respectively that we constructed in Section~\ref{benchmark}.}
\label{training_in_domian}
\end{figure*}

\begin{figure*}[t]
    \centering
    \subfigure[$0\%$ on Flicker30k]{
	\includegraphics[width=1in]{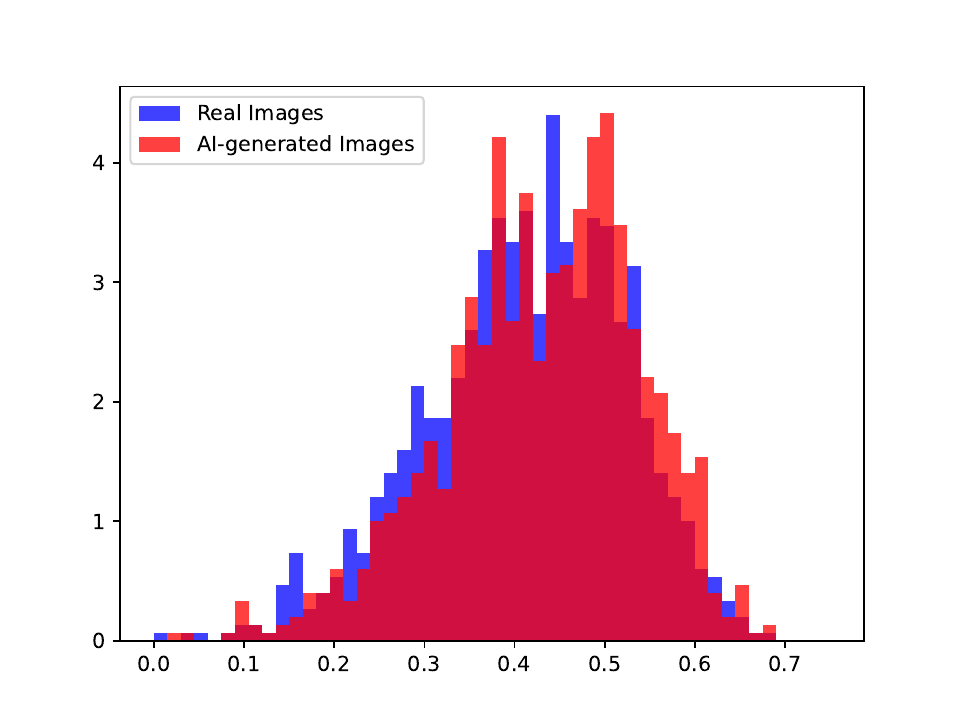}
	\label{ratio_r}
    }\quad
    \subfigure[$20\%$ on Flicker30k]{
	\includegraphics[width=1in]{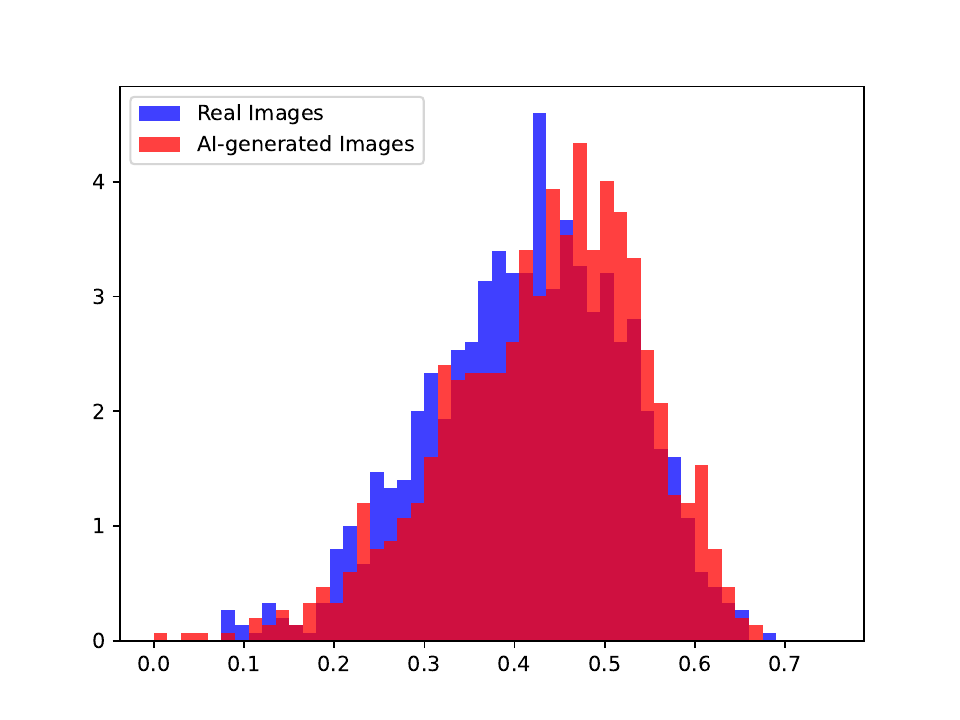}
	\label{ratio_r}
    }\quad
    \subfigure[$40\%$ on Flicker30k]{
	\includegraphics[width=1in]{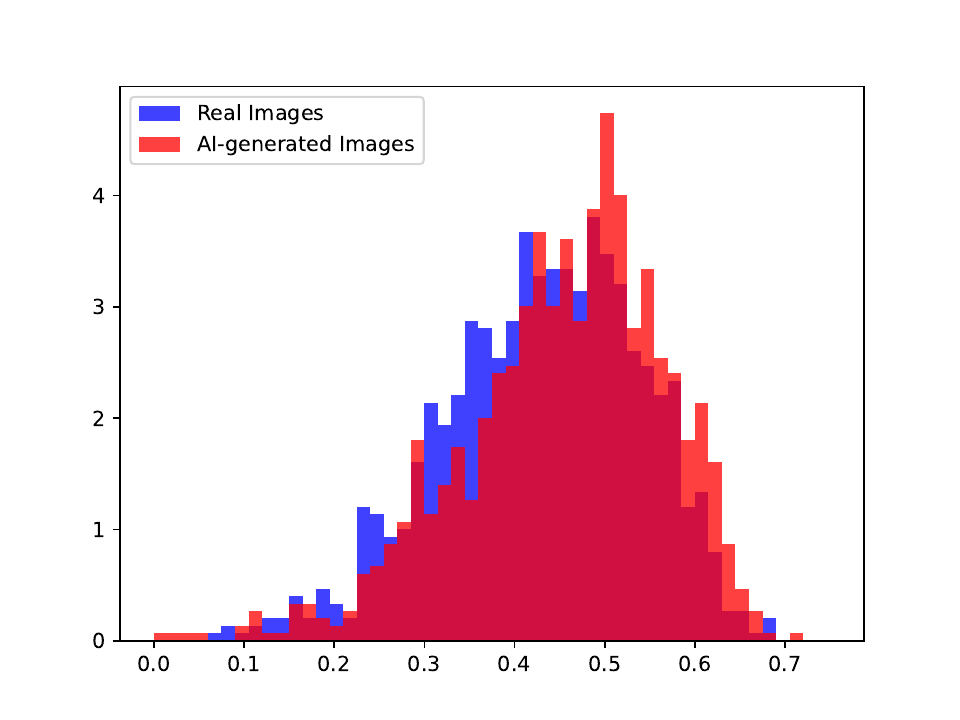}
	\label{ratio_r}
    }\quad
    \subfigure[$60\%$ on Flicker30k]{
	\includegraphics[width=1in]{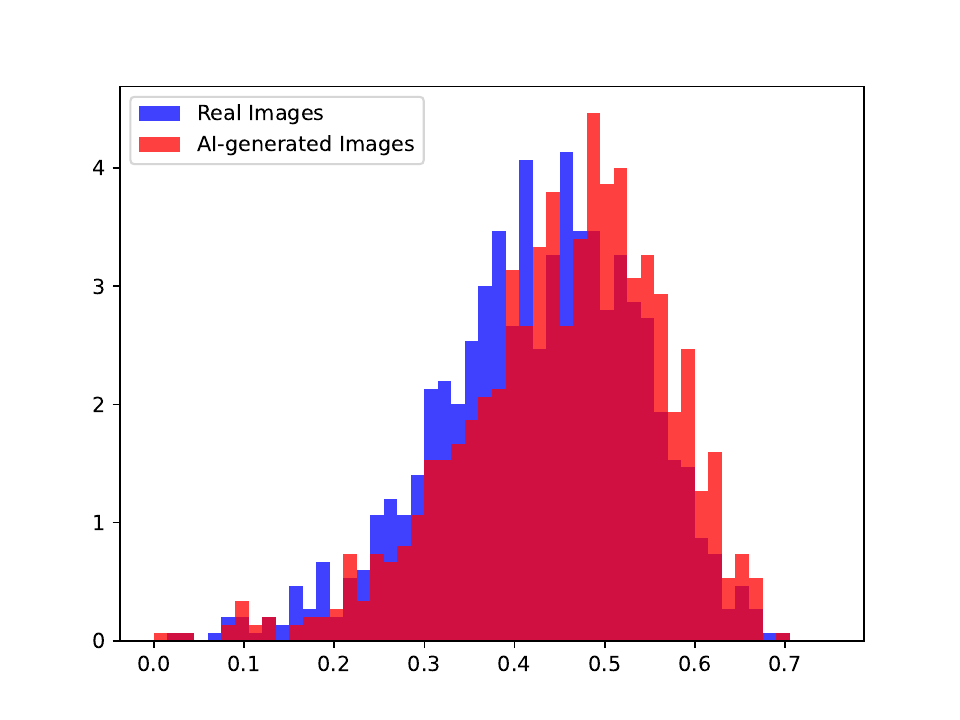}
	\label{ratio_r}
    }\quad
    \subfigure[$80\%$ on Flicker30k]{
	\includegraphics[width=1in]{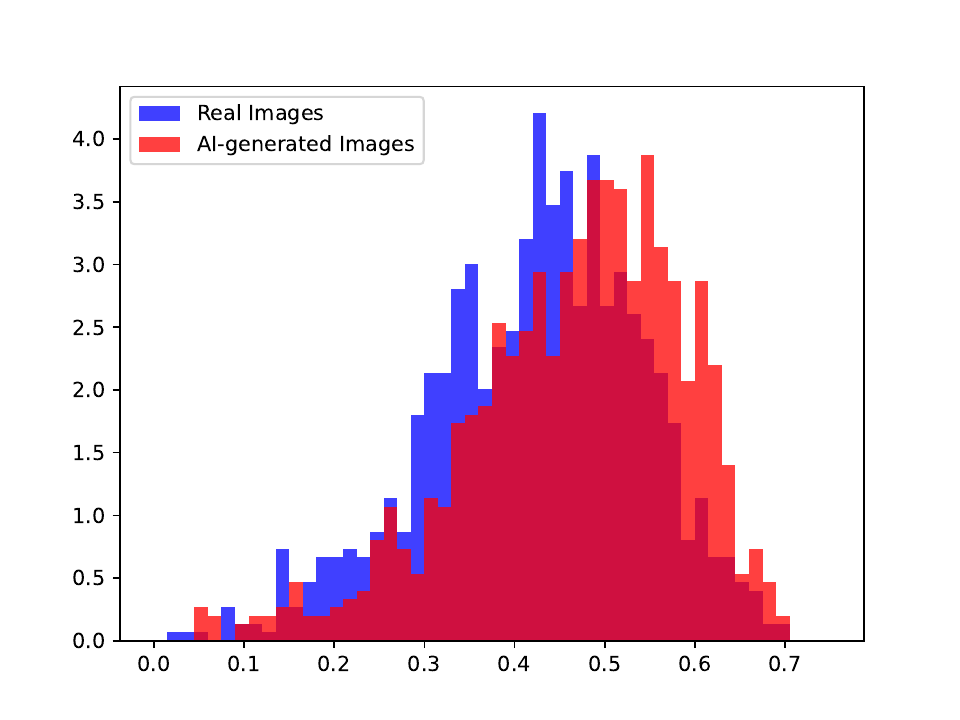}
	\label{ratio_r}
    }\quad
        \subfigure[$100\%$ on Flicker30k]{
	\includegraphics[width=1in]{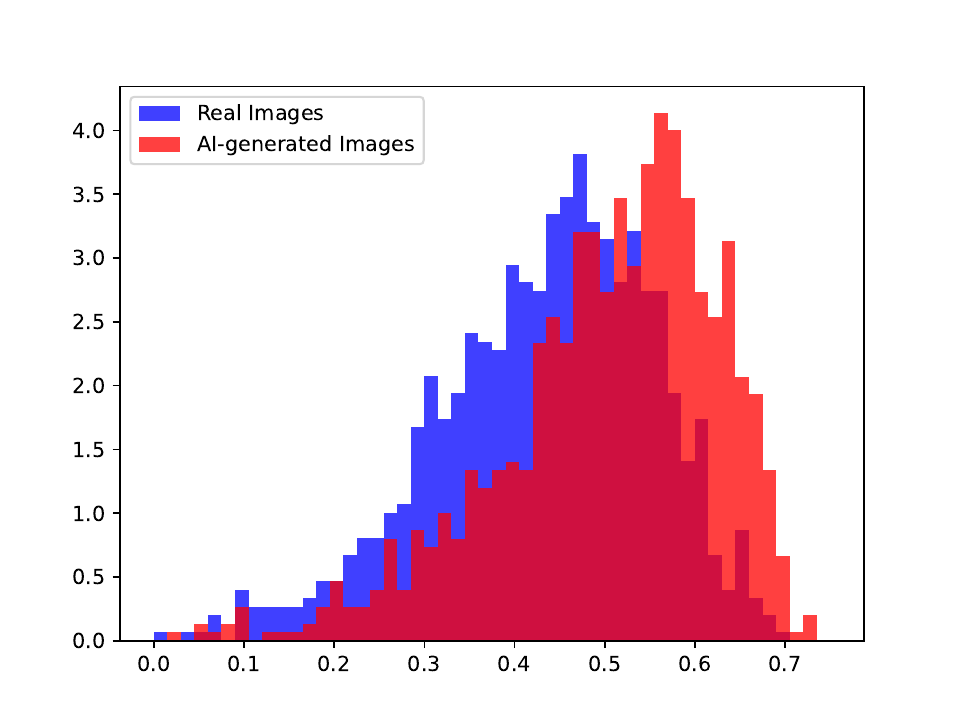}
	\label{ratio_r}
    }
        \subfigure[$0\%$ on MSCOCO]{
	\includegraphics[width=1in]{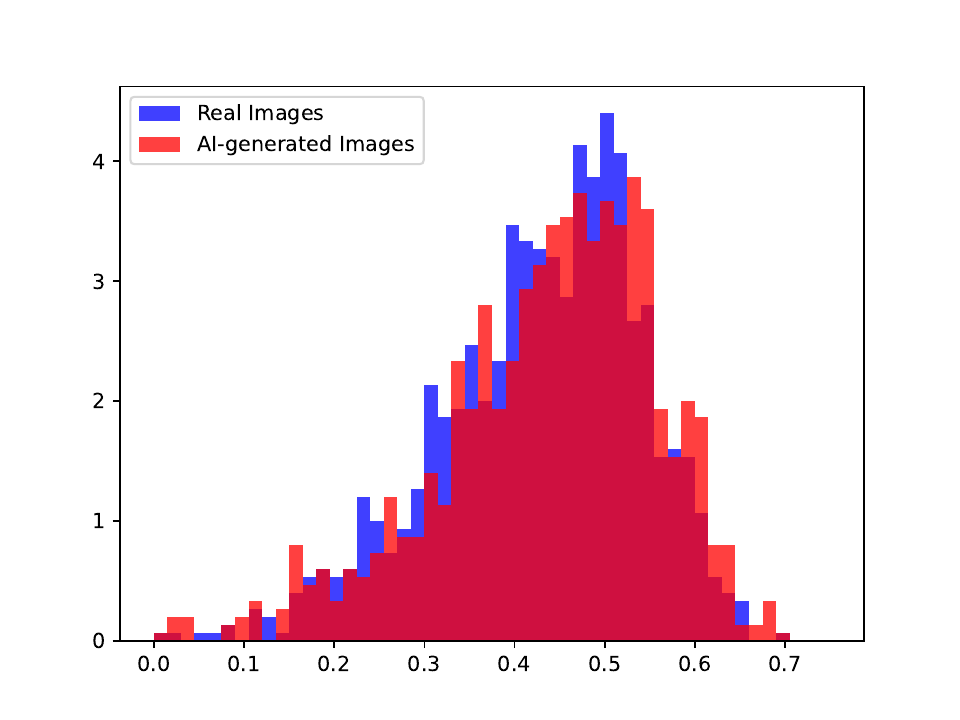}
	\label{ratio_r}
    }\quad
    \subfigure[$20\%$ on MSCOCO]{
	\includegraphics[width=1in]{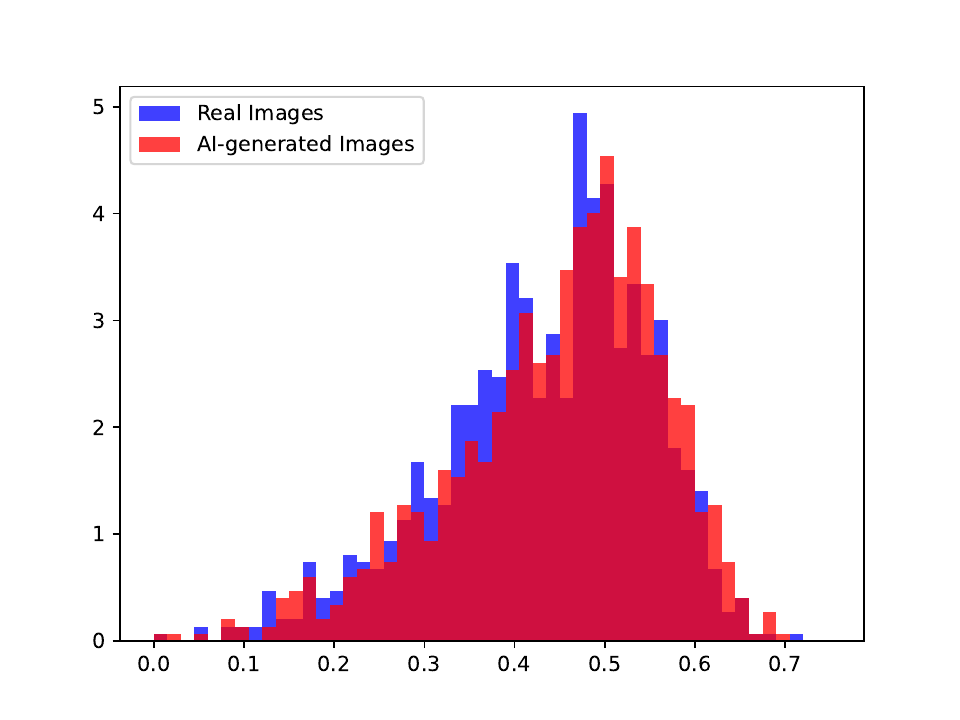}
	\label{ratio_r}
    }\quad
    \subfigure[$40\%$ on MSCOCO]{
	\includegraphics[width=1in]{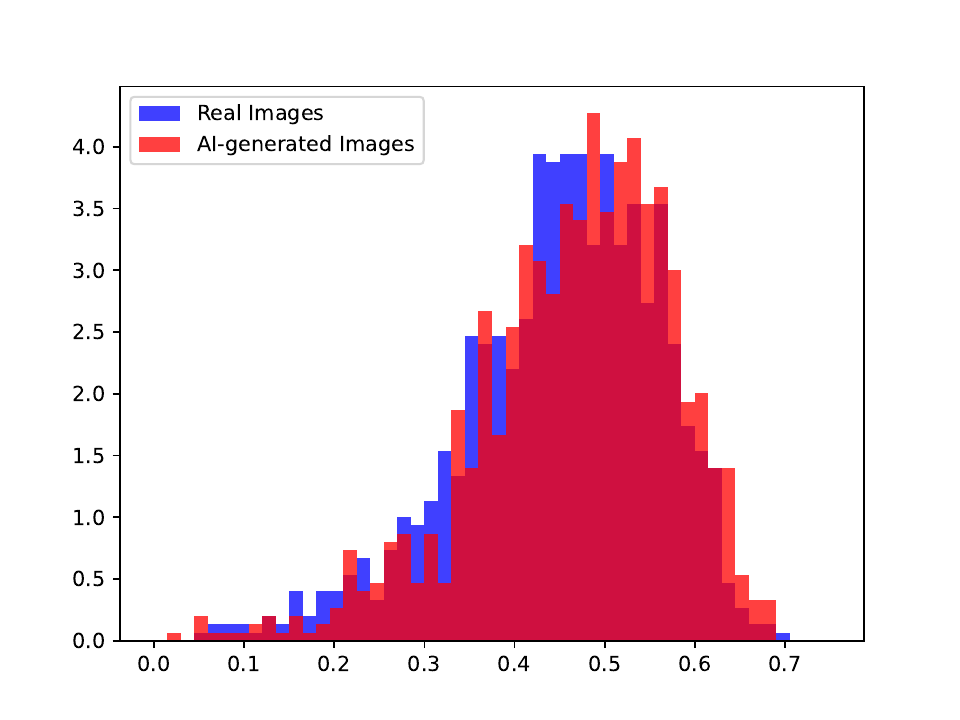}
	\label{ratio_r}
    }\quad
    \subfigure[$60\%$ on MSCOCO]{
	\includegraphics[width=1in]{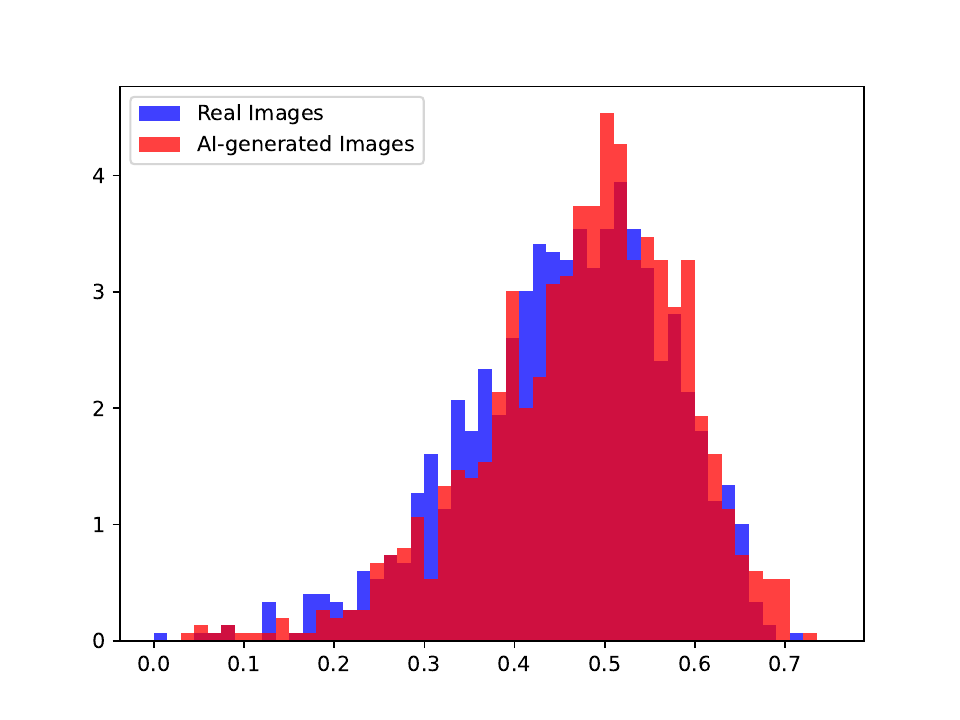}
	\label{ratio_r}
    }\quad
    \subfigure[$80\%$ on MSCOCO]{
	\includegraphics[width=1in]{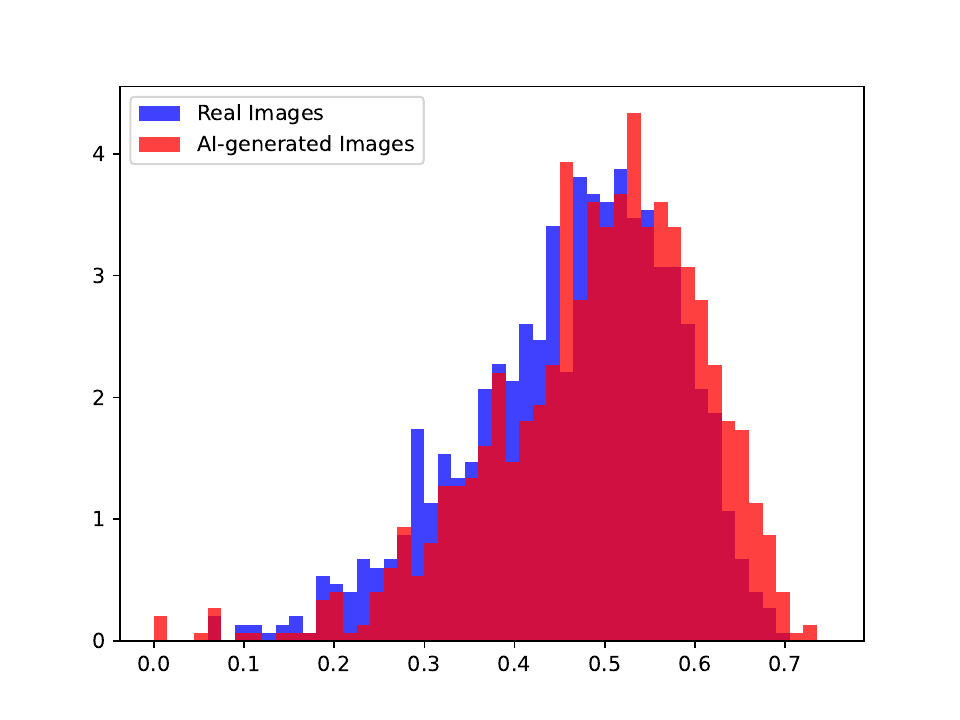}
	\label{ratio_r}
    }\quad
        \subfigure[$100\%$ on MSCOCO]{
	\includegraphics[width=1in]{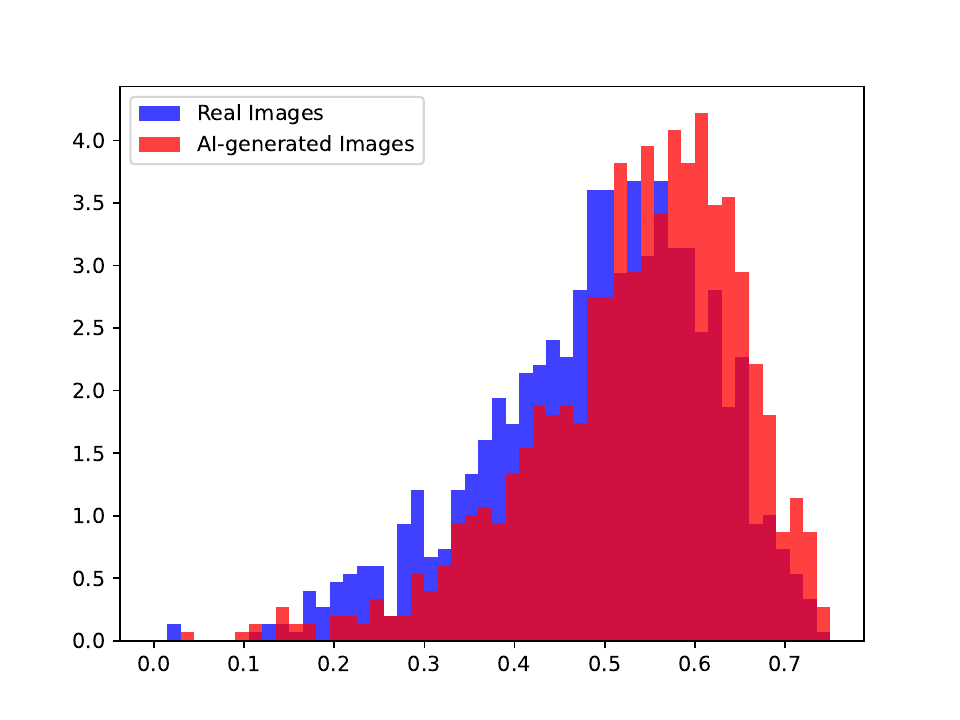}
	\label{ratio_r}
    }
\caption{Distribution of the caption-image relevance scores estimated by retrieval models that are trained on the datasets mixed with different ratios of AI-generated images. Flicker30k is in-domain and MSCOCO is out-of-domain.}
\label{training_distribution}
\end{figure*}

Figure~\ref{training_in_domian} illustrates that as the ratio increases, the ranking disparity between generated images and real images widens, with the retrieval model exhibiting a greater inclination to rank generated images higher ($\rm{Relative} \triangle$ decreasing). Meanwhile, Figure~\ref{training_distribution} demonstrates that with an increasing ratio, the discrepancy in score distribution between generated images and real images increases, and the scores of AI-generated images gradually become greater and greater than those of real images. In both in-domain and out-of-domain settings, the following conclusions can be drawn from these results: (1) Retrieval models trained on the datasets mixed with AI-generated images exhibit more serious invisible relevance bias. (2) The invisible relevance bias tends to become more serious as the ratio of AI-generated images in the training data increases. (3) As the ratio of AI-generated images in the training data increases, the retrieval performance on real images is gradually damaged.

\section{Our Method in debiasing} \label{debias_method}
This section proposes an effective method to alleviate the bias.


\subsection{Design of Debiasing Method}

Our method alleviates the retrieval model's preference bias for generated images by measuring and optimizing the additional relevance score of generated images in training. Given a caption-image pair $(C_i, I^r_i)$ in the training set, the training process for both dual-encoder and fusion-encoder text-image retrieval models can be formulated as estimating the relevance score $s$ between $C_i$ and $I^r_i$, and using contrastive loss or regression loss as the optimization objective to adjust $s$, which can be described as:
\begin{equation}
  \begin{aligned}
     s = R(C_i,I^r_i;\theta), \quad 
    \theta_{\text{optimal}} = \arg \min_\theta \mathcal{L}(s, y; \theta),
  \end{aligned}
\nonumber
\end{equation}
in which $R(\cdot;\theta)$ is the relevance scoring function of the retrieval model such as cosine similarity between representations in the dual-encoder model and neural networks in the fusion-encoder model, $\theta$ is the set of parameters of the model, $\mathcal{L}$ is the loss function such as contrastive loss or regression loss, $y$ is the label. In our method, for each real image $I^r_i$ in the training data, we use the method in Section~\ref{benchmark} to generate its corresponding AI-generated image $I^g_i$. Then we can get the caption-real-AI triple as $(C_i, I^r_i, I^g_i)$, in which $C_i$ is the paired caption for the real image $I^r_i$. We introduce the contrastive loss to get the difference in relevance scores between $I^r_i$ and $I^g_i$ with respect to the caption $C_i$ as:
\begin{equation}
  \begin{aligned}
     \triangle s(I^g_i,I^r_i) &= R(C_i,I_{i}^{g};\theta) - R(C_i,I^r_i;\theta),
  \end{aligned}
\nonumber
\end{equation}
which can measure the additional invisible relevance introduced by the AI-generated image $I^g_i$ for $C_i$ compared with the real image $I^r_i$. Therefore, this can be used as a part of the optimization objective to mitigate the invisible relevance bias. It is because minimizing the difference between $R(C_i, I_{i}^{g};\theta)$ and $R(C_i, I^r_i;\theta)$ in training can make the retrieval model eliminate the additional score estimated for $I^g_i$. In all triples in the training data, we consider the caption-real-AI triples $(C_i, I^r_i, I^g_i)$ whose $\triangle s(I^r_i, I^g_i)$ is greater than $0$ and perform Bernoulli sampling from these triples with probability $\beta$ to get the target triple set $\mathbb{B}$ for debiasing optimization as:
\begin{equation}
\mathbb{B}= \{ (C_i,I^r_i,I^g_i) \mid (C_i,I^r_i,I^g_i) \sim \text{Bernoulli}(\beta) \cdot \mathbb{I}(\triangle s(I^g_i,I^r_i) > 0) \}.
\nonumber
\end{equation}
The reason why we only sample triples with probability $\beta$ is to adjust the tolerance of the retrieval models to the AI-generated images. The higher the probability $\beta$, the more likely the retrieval models are to rank the AI-generated images to a lower position. The total optimization objective in training is:
\begin{equation}
  \begin{aligned}
    \theta_{\text{optimal}} = \arg \min_\theta (\sum_{s_i,y_i \in \mathbb{A}}\mathcal{L}(s_i, y_i; \theta) + \sum_{I^r_i,I^g_i \in \mathbb{B}} \triangle s(I^g_i,I^r_i)),
  \end{aligned}
\nonumber
\end{equation}
in which $\mathbb{A}$ is the set of all samples in the training data. For the sample $i$, $s_i$ is the estimated score and $y_i$ is the label.

\subsection{Evaluation of Debiasing}

 \begin{table*}[t]
  \caption{Performance of the retrieval models on the benchmark we constructed consisting of both real and AI-generated images with different sampling probability $\beta$ in our debiasing method. $\rm{Relative \triangle} > 0$ means retrieval models rank real images higher than AI-generated images, $\rm{Relative \triangle} < 0$ means retrieval models rank AI-generated images higher than real images. When $\rm{Relative \triangle} < 0$, the absolute value of $\rm{Relative \triangle}$ indicates the value of this bias.}
  \label{debias}
\renewcommand\arraystretch{1.0}
\setlength\tabcolsep{3pt}
\scalebox{0.75}{
\begin{tabular}{lcllllllcllllll}
\toprule
                                    & \multicolumn{7}{c}{Flicker30k+AI (In-domain)} & \multicolumn{7}{c}{MSCOCO+AI (Out-of-domain)}     \\ \hline
                                    & w/o debias     & $\beta=50\%$   & $\beta=60\%$  & $\beta=70\%$  & $\beta=80\%$ & $\beta=90\%$   & $\beta=100\%$  & w/o debias    & $\beta=50\%$   & $\beta=60\%$  & $\beta=70\%$  & $\beta=80\%$ & $\beta=90\%$   & $\beta=100\%$   \\
$\rm{Relative} \triangle$ on NDCG@1 & -10.35  & -1.406 & 31.42 & 62.77 & 91.71 & 112.06 & 129.20 & -13.53 & -1.384 & 45.35 & 80.67 & 114.63 & 140.23 & 154.43 \\
$\rm{Relative} \triangle$ on NDCG@3 & -4.31   & -0.656 & 15.08 & 32.85 & 50.08 & 65.44  & 77.31  & -3.64  & -0.354 & 27.61 & 53.11 & 78.92  & 101.67 & 114.31 \\
$\rm{Relative} \triangle$ on NDCG@5 & -4.37   & -0.876 & 13.13 & 27.84 & 42.28 & 55.68  & 65.31  & -2.22  & -0.214 & 23.47 & 46.42 & 69.62  & 90.28  & 102.89 \\
NDCG@1 on only real images          & 30.57   & 33.44  & 33.15 & 33.26 & 33.12 & 33.09  & 33.20  & 18.50  & 21.09  & 21.48 & 21.32 & 20.52  & 20.43  & 20.01  \\
NDCG@1 on only real images          & 37.95   & 40.44  & 40.32 & 40.53 & 40.38 & 40.13  & 40.31  & 25.66  & 28.92  & 29.05 & 28.78 & 28.32  & 28.05  & 27.65  \\
NDCG@5 on only real images          & 39.78   & 42.29  & 41.98 & 42.18 & 42.05 & 41.93  & 42.10  & 28.28  & 31.55  & 31.69 & 31.43 & 30.97  & 30.56  & 30.02  \\ \toprule
\end{tabular}
}
\end{table*}

\begin{figure*}[t]
    \centering
    \subfigure[$\beta=50\%$ on Flicker30k]{
	\includegraphics[width=1in]{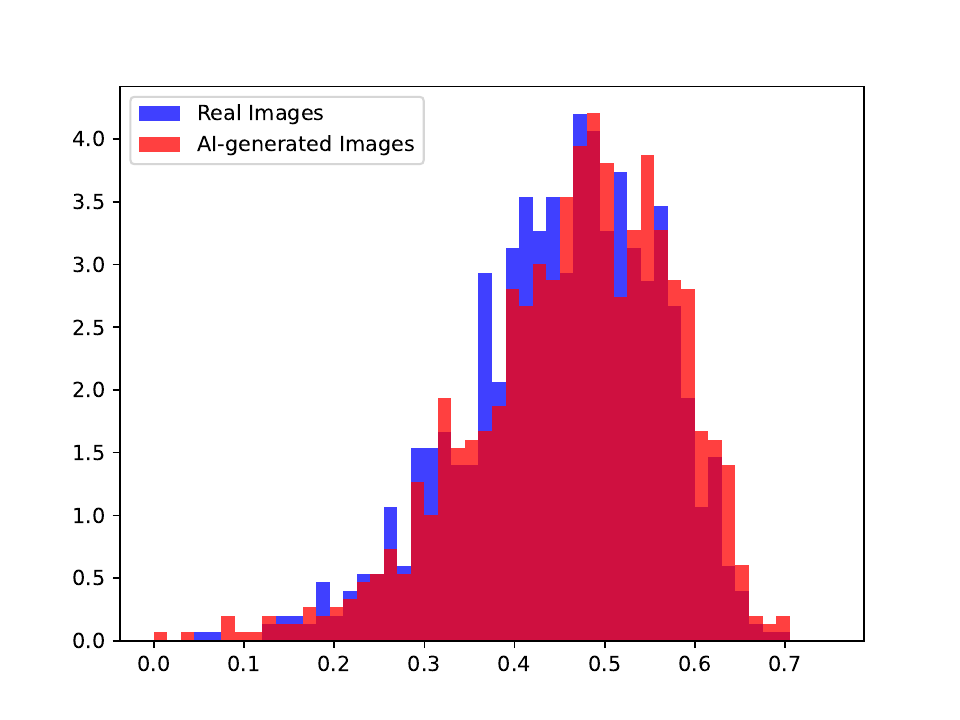}
	\label{ratio_r}
    }
    \subfigure[$\beta=60\%$ on Flicker30k]{
	\includegraphics[width=1in]{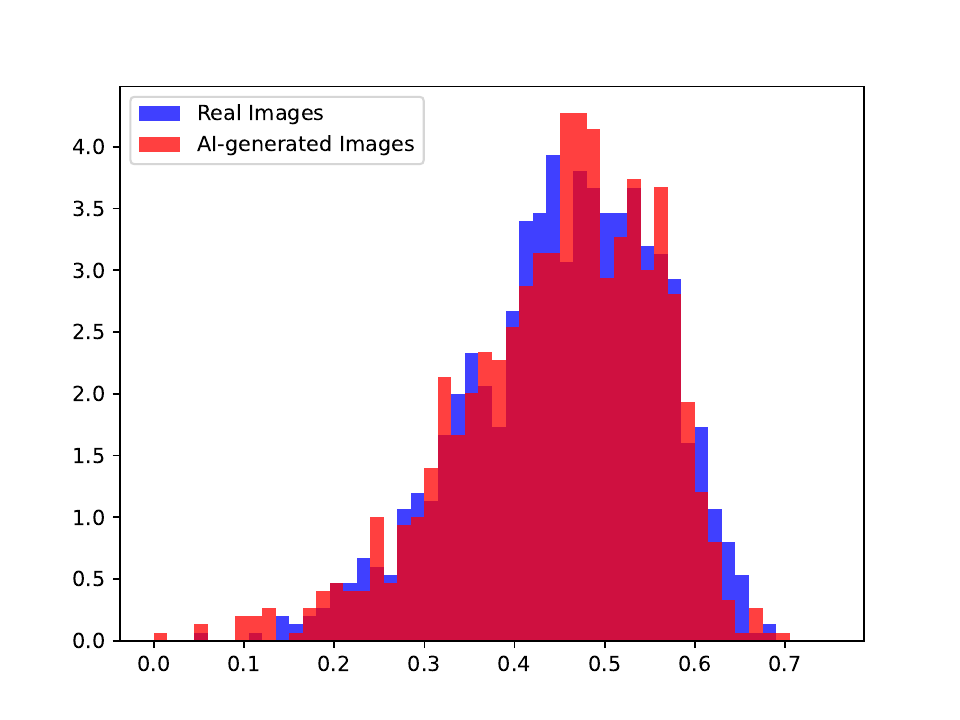}
	\label{ratio_r}
    }
    \subfigure[$\beta=70\%$ on Flicker30k]{
	\includegraphics[width=1in]{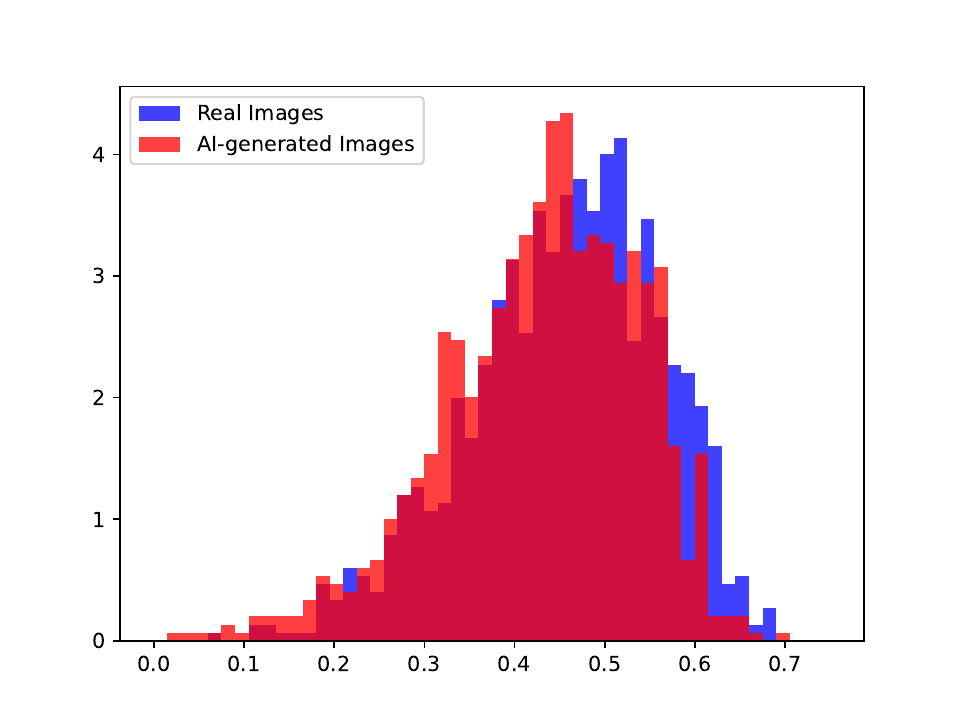}
	\label{ratio_r}
    }
    \scalebox{0.975}{\subfigure[$\beta=80\%$ on Flicker30k]{
	\includegraphics[width=1in]{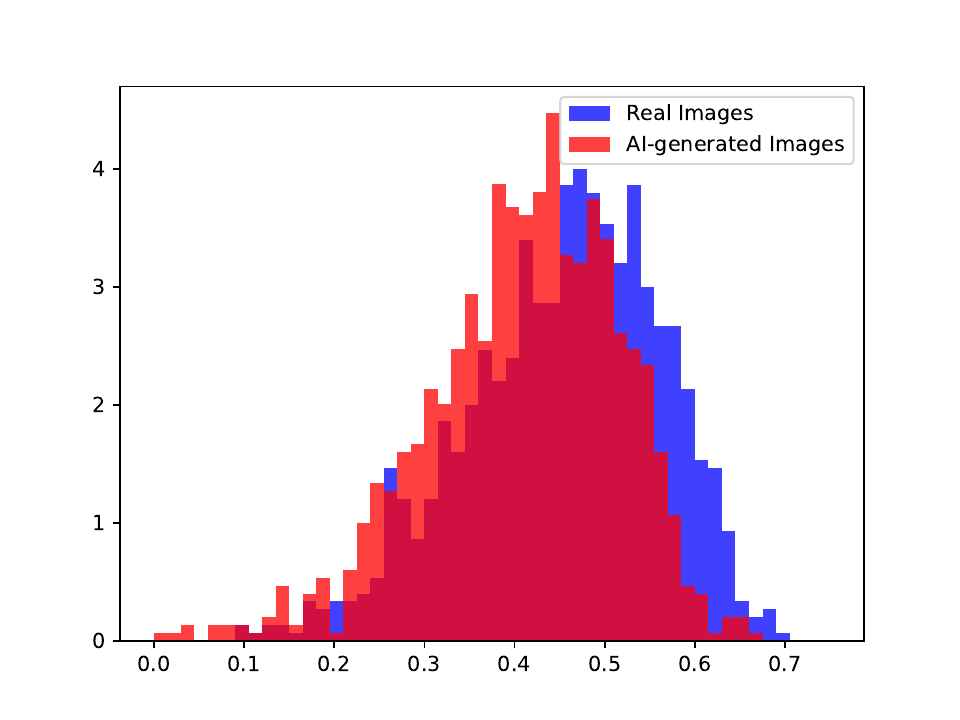}
	\label{ratio_r}
    }}
    \scalebox{0.975}{
    \subfigure[$\beta=90\%$ on Flicker30k]{
	\includegraphics[width=1in]{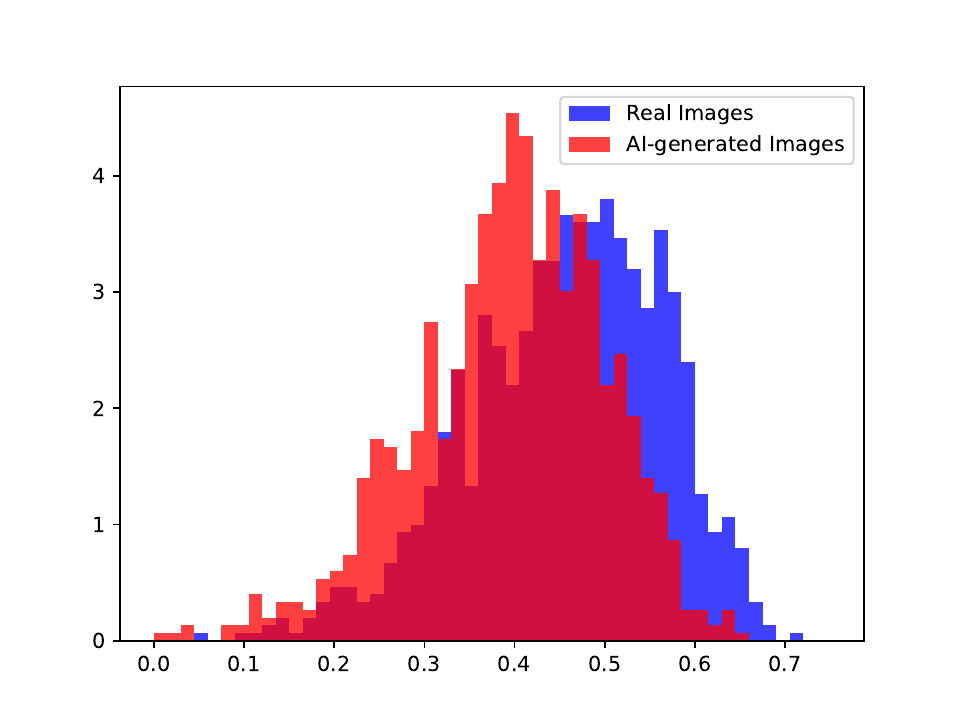}
	\label{ratio_r}
    }}
    \scalebox{0.975}{\subfigure[$\beta=100\%$ on Flicker30k]{
	\includegraphics[width=1in]{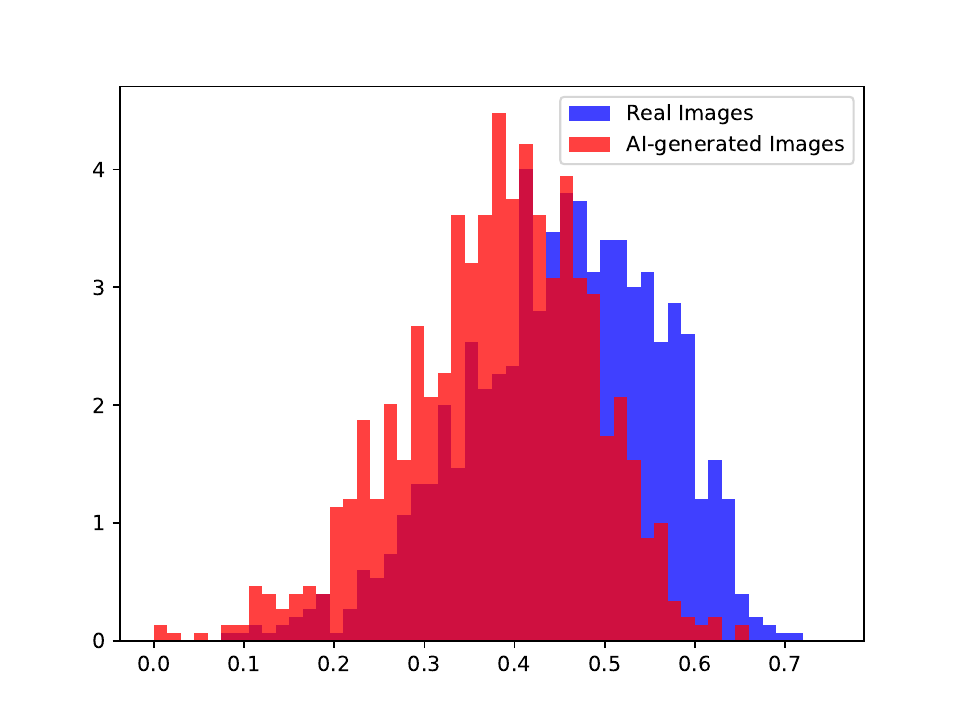}
	\label{ratio_r}
    }}
        \subfigure[$\beta=50\%$ on MSCOCO]{
	\includegraphics[width=1in]{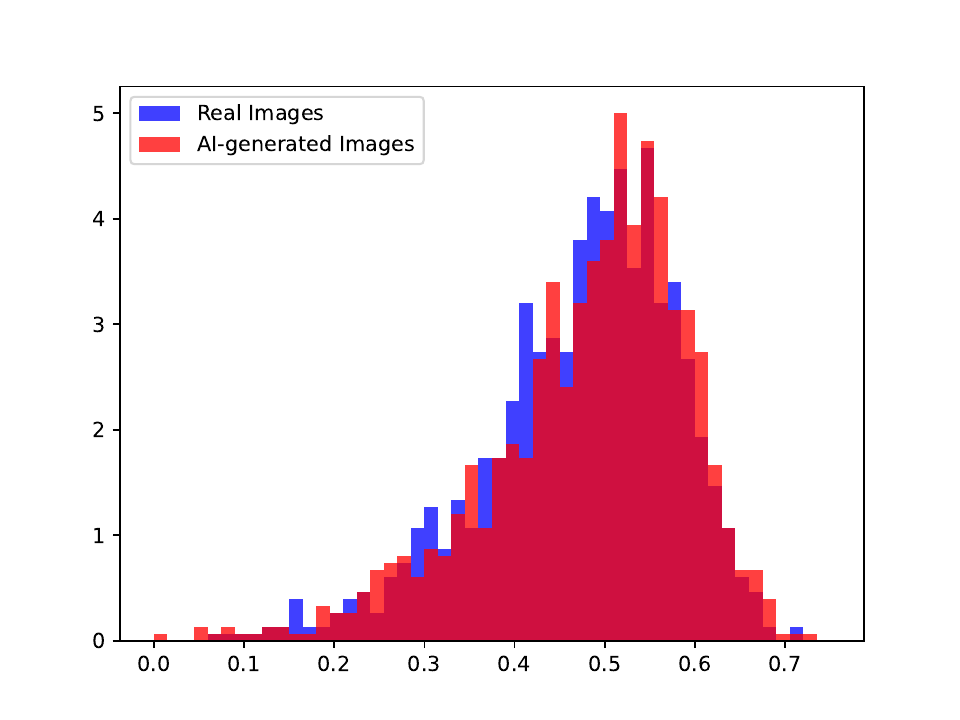}
	\label{ratio_r}
    }
    \subfigure[$\beta=60\%$ on MSCOCO]{
	\includegraphics[width=1in]{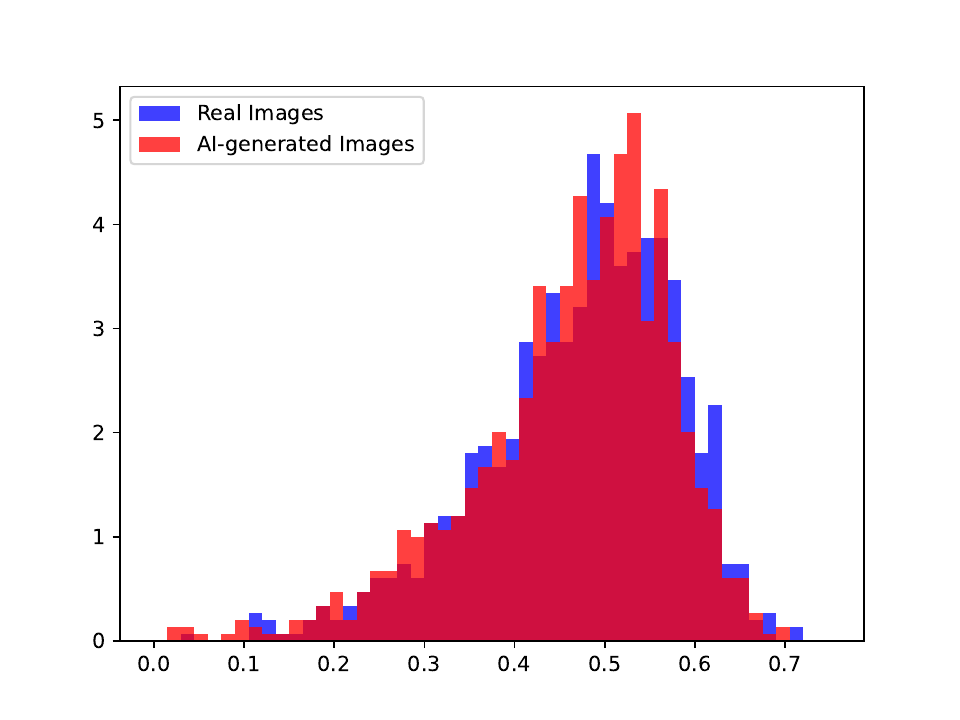}
	\label{ratio_r}
    }
    \subfigure[$\beta=70\%$ on MSCOCO]{
	\includegraphics[width=1in]{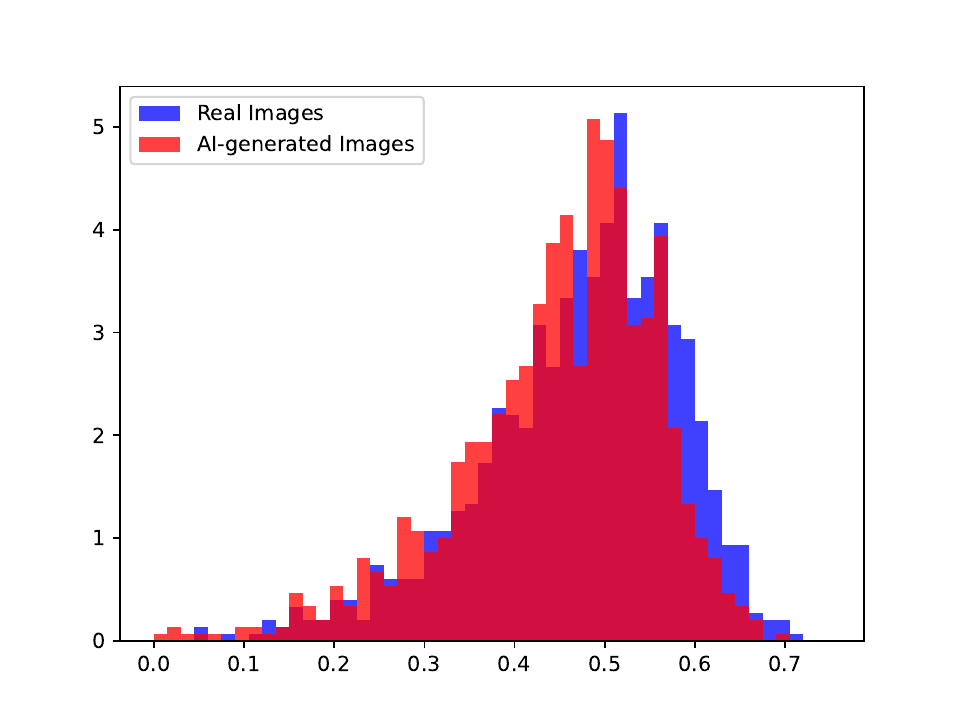}
	\label{ratio_r}
    }
    \subfigure[$\beta=80\%$ on MSCOCO]{
	\includegraphics[width=1in]{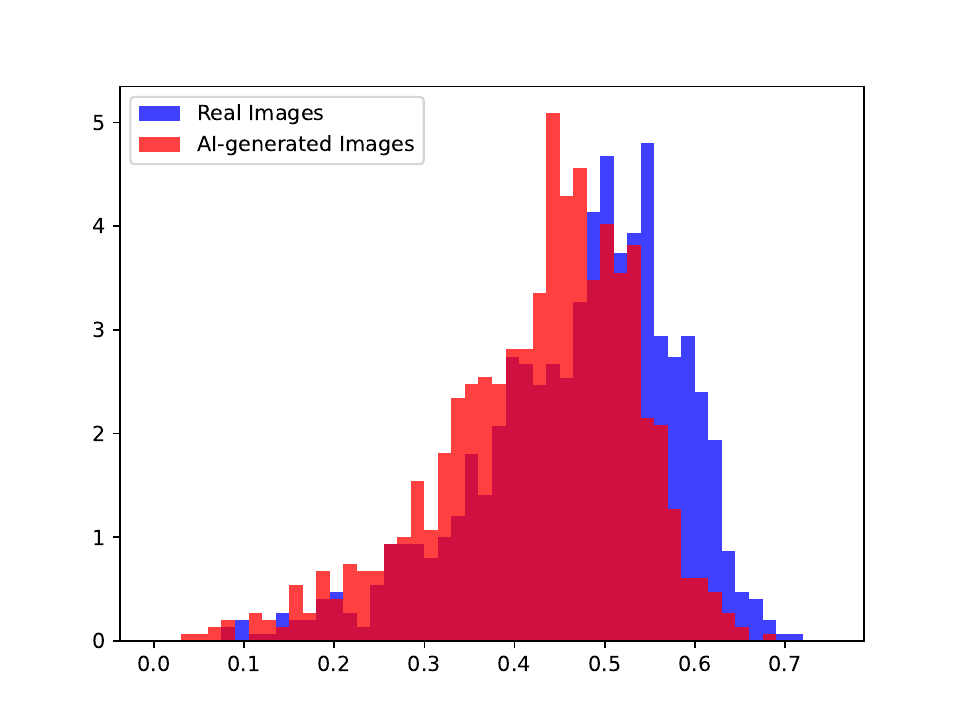}
	\label{ratio_r}
    }
    \subfigure[$\beta=90\%$ on MSCOCO]{
	\includegraphics[width=1in]{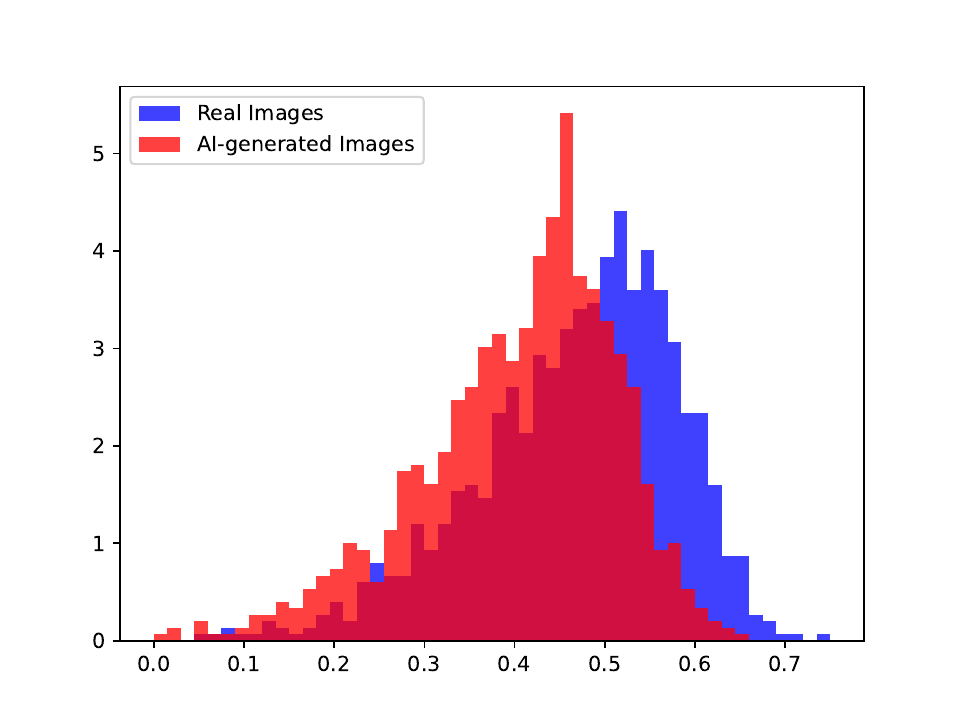}
	\label{ratio_r}
    }
        \subfigure[$\beta=100\%$ on MSCOCO]{
	\includegraphics[width=1in]{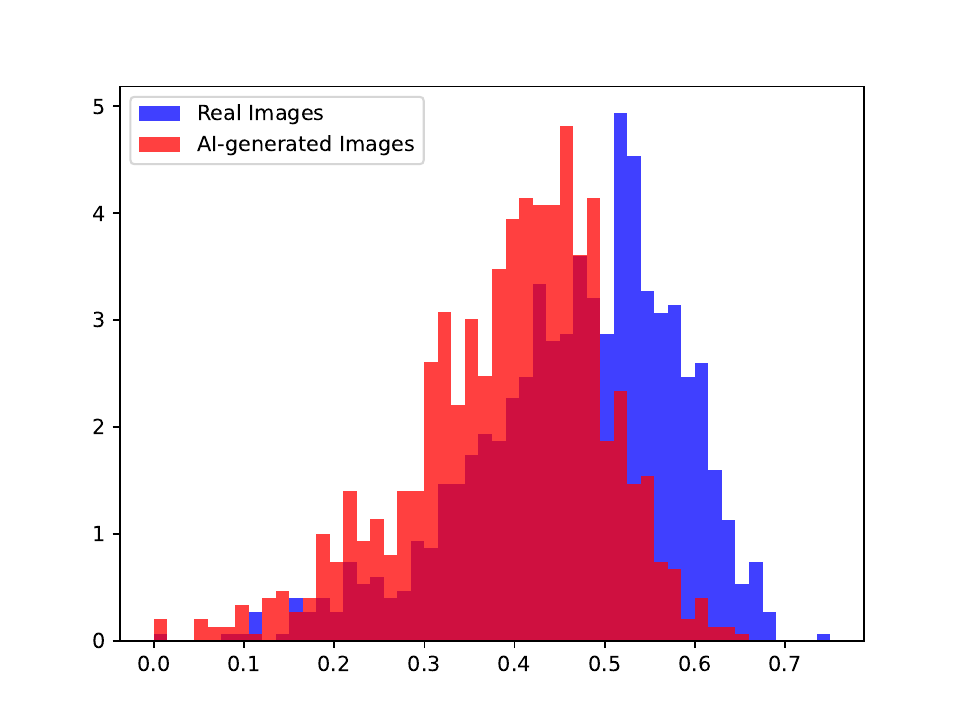}
	\label{ratio_r}
    }
\caption{Distribution of the caption-image relevance scores estimated by retrieval models with different sampling probability $\beta$ in our debiasing method. Flicker30k is in-domain and MSCOCO is out-of-domain.}
\label{debias_distribution}
\end{figure*}

Evaluation of our debiasing method focuses on: (1) How it affects the ranking difference between real and AI-generated images. (2) How it affects the distribution of caption-image relevance scores.


Table~\ref{debias} shows how the $\rm{Relative \triangle}$ and retrieval performance changes with the sampling probability $\beta$. The results indicate that: (1) Our method not only effectively alleviates the retrieval model's preference for AI-generated images, but also makes real images ranked significantly higher than AI-generated images. (2) As the sampling probability $\beta$ increases, real images are ranked higher and higher than AI-generated images. (3) When $\beta$ is $0.5$, retrieval models can achieve a fair ranking between real and generated images with little bias. (4) Our method improves the retrieval performance. It is because, in our method, AI-generated images can be seen as the hard negatives in training, which enhances the ability of the retrieval model to distinguish the images with very similar semantics.

Figure~\ref{debias_distribution} shows the distribution of the caption-image relevance scores estimated by retrieval models with different sampling probability $\beta$. The results indicate that our method effectively reduces the relevance scores between captions and generated images estimated by the retrieval model. With the increase in sampling probability ($\beta$), the disparity in score distribution between generated images and real images expands, and the scores of real images gradually become greater and greater than those of generated images.



\section{Causes of Invisible Relevance Bias} \label{cause}

In this section, we use the debiased model proposed in Section~\ref{debias_method} to reversely analyze the causes of the invisible relevance bias. Specifically, we extract the transformations performed by the debiased retriever on AI-generated images. The reverse process of these transformations can be seen as the cause of invisible relevance bias. We also design the experiments to support this point of view. 

\subsection{Transformations in Debiasing}
To simplify the analysis process, we focus on the dual-encoder retrieval model that estimates the relevance score by computing vector similarity between text and image representations. For the same text, image representation can directly affect the estimation of the relevance, so we use image representation as the main object in our analysis, which can help us to find the most direct causes of invisible relevance bias. Specifically, we analyze the difference between image representations encoded by the original retriever and debiased retriever and extract the transformations of debiasing from this difference. Then, we reverse these transformations to explore the causes of the invisible relevance bias. 

Given the image encoder $v(\cdot;\theta)$ of original retriever, the image encoder $v^{d}(\cdot;\theta')$ of our debiased retriever, and an AI-generated image $I^g$, we can get the representations of $I^g$ encoded by $v(\cdot;\theta)$ and $v^{d}(\cdot;\theta')$ respectively as $ \boldsymbol{r} = v(I^g;\theta), \boldsymbol{r^d} = v^{d}(I^g;\theta').$
For the normalized text representation $\boldsymbol{t}=[t_1,t_2,...,t_n]$, image representation $\boldsymbol{r}=[r_1,r_2,...,r_n]$ from the original retriever and image representation $\boldsymbol{r^d}=[r^d_1,r^d_2,...,r^d_n]$ from the debiased retriever, the relevance $score( \boldsymbol{t}, \boldsymbol{r})$ and $score(\boldsymbol{t},\boldsymbol{r^d})$ can be calculated as:
\begin{equation}
  \begin{aligned}
    score(\boldsymbol{t},\boldsymbol{r}) &= {t_1}{r_1} + {t_2}{r_2} + {t_3}{r_3} + ... + {t_n}{r_n}, \\
      score(\boldsymbol{t},\boldsymbol{r^d}) &= {t_1}{r^d_1} + {t_2}{r^d_2} + {t_3}{r^d_3} + ... + {t_n}{r^d_n}.   
  \end{aligned}
\nonumber
\end{equation}
Therefore, the adjustment of relevance score between text $t$ and AI-generated image $I^g$ in the debiased model is essentially changing the values of each element ($r_i$) in $\boldsymbol{r}$, which can be described as:
\begin{equation}
  \begin{aligned}
      score(\boldsymbol{t},\boldsymbol{r^d}) &= {t_1}{r^d_1} + {t_2}{r^d_2} + ... + {t_n}{r^d_n} \\
      &=  {t_1}{(r_1+ \triangle r_1)} + {t_2}{(r_2 + \triangle r_2)} + ... + {t_n}{(r_n + \triangle r_n)}. 
  \end{aligned}
\nonumber
\end{equation}
The transformations $\boldsymbol{p}$ in the debiased model can be represented by a vector with the same dimensions as $r$ and $r^d$:
\begin{equation}
  \begin{aligned}
      \boldsymbol{p} &= [\triangle r_1,\triangle r_2,...,\triangle r_n] = [p_1,p_2,...,p_n] ,
  \end{aligned}
\nonumber
\end{equation}
Then, we perform two-dimensional visualization of the $\boldsymbol{r}$, $\boldsymbol{r^d}$ and $\boldsymbol{p}$ of all images in datasets to try to find the patterns from them. The T-SNE visualization in Figure~\ref{tsne_casue} shows that compared with the scattered image representations, the transformations vector $\boldsymbol{p}$ shows an obvious aggregation phenomenon. This indicates that there is consistency in the transformations performed by the debiased retriever on AI-generated images with very different semantics. 


\subsection{Reversing the Transformations}
Debiased retriever modifies each element ($r_i$) of the representation $\boldsymbol{r}$ from the original retriever according to the value of the corresponding element $p_i$ in transformations vector $\boldsymbol{p}$ and gets the debiased representation $\boldsymbol{r^d}$, which can be described as:
\begin{equation}
  \begin{aligned}
      r^d_i = r_i + p_i, r_i \in \boldsymbol{r}, p_i \in \boldsymbol{p}, r^d_i \in \boldsymbol{r^d}.
  \end{aligned}
\nonumber
\end{equation}
\begin{figure}[t]
\centering
	\includegraphics[width=0.5\linewidth, height=1.5in]{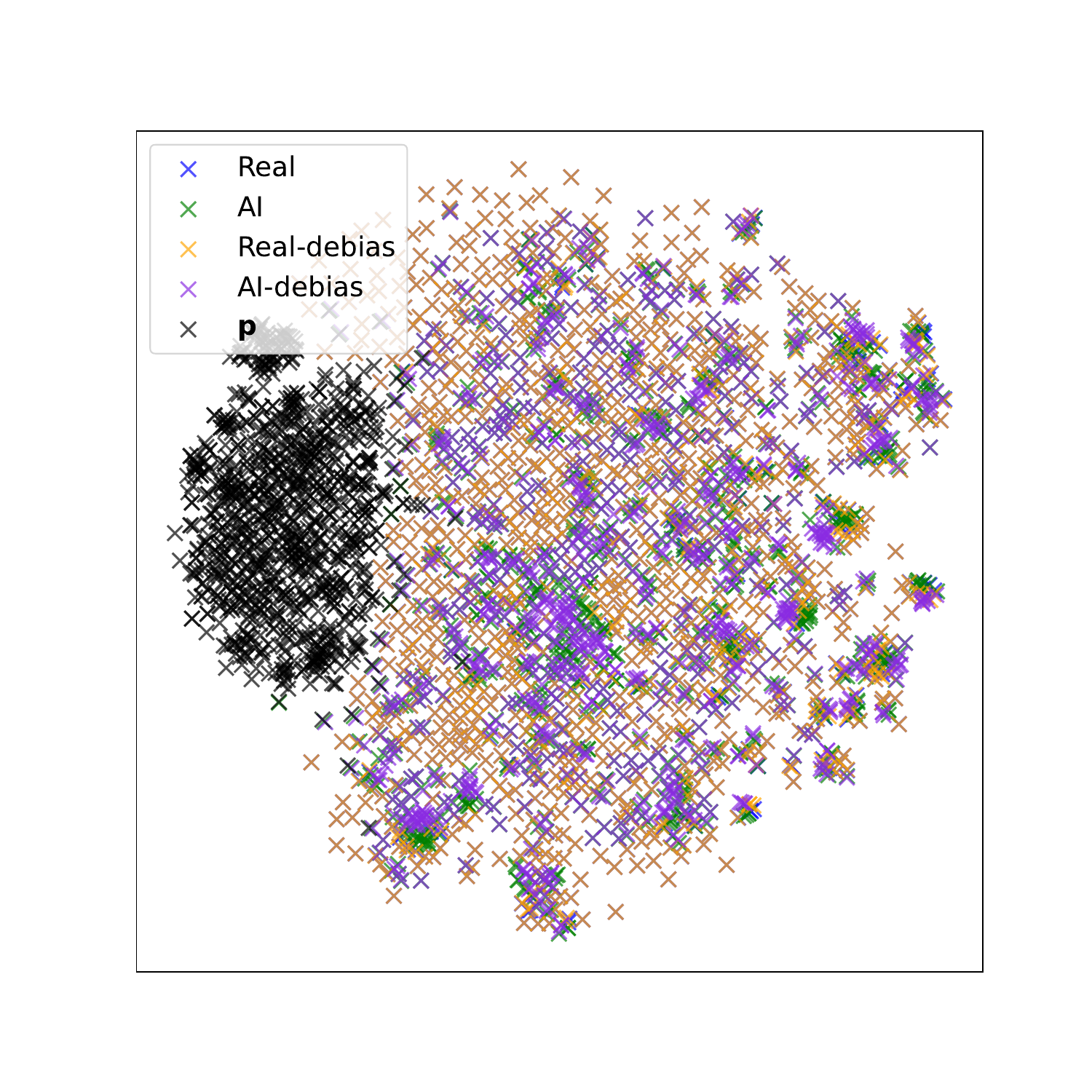}
\caption{T-SNE visualization of image representations and transformations vector $\boldsymbol{p}$.}
        \label{tsne_casue}
\end{figure}
We can reverse this process to get the causes of the invisible relevance bias. This is because the transformation made by a debiased retriever to make a biased AI-generated image become unbiased is exactly the difference between the biased AI-generated images and the real images. That is, the reason why an unbiased representation $\boldsymbol{r^d}$ becomes the representation $\boldsymbol{r}$ with bias is that the reverse transformation ($-p_i$) is done on $r^d_i$ ($r_i = r^d_i - p_i$). Therefore, we conclude that an unbiased representaion $\boldsymbol{r^d}$ becomes the representaion $\boldsymbol{r}$ with bias because $\boldsymbol{r^d}$ is added bitwisely by a vector $\boldsymbol{-p}=[-p_1,-p_2,-p_3,...,-p_n]$. Combining the phenomenon shown in Figure~\ref{tsne_casue} that there is consistency in the transformations vector $\boldsymbol{p}$ on AI-generated images with very different semantics, we can get the causes of invisible relevance bias in AI-generated images: \textbf{AI-generated images cause the image encoder in the retriever to embed additional information to their representations.} This additional information is the direct cause of invisible relevance bias and can be obtained by the difference in image representations between original and debiased retrievers. This information has the following three characteristics:  \textbf{(1)} This information cannot be reflected in a visible way, but can only be embedded by neural network-based models. \textbf{(2)} When this information is embedded into the representation of the image, it can amplify the query-image relevance to produce a higher score. \textbf{(3)} This information has a certain consistency in AI-generated images with different semantics. This information may be like the watermark that is universal information for the image generation model and can be expressed by neural network visual models such as image encoders. We design the experiments to support these three points.

\subsection{Supporting Experiments}
\begin{table}[t]
  \caption{Effect of reverse transformations vector $\boldsymbol{-p}$ on test set of Flicker30k. The retrieval model is VSE trained on Flicker30k without any debiasing training.}
\setlength\tabcolsep{4pt}
\renewcommand\arraystretch{1.05}
\centering
\scalebox{0.75}{
\begin{tabular}{lcccccc}
\toprule
  & \multicolumn{6}{c}{Relative $\triangle$ on }     \\ 
 & NDCG@1 & NDCG@3 & NDCG@5 & R@1    & R@3    & R@5   \\ \hline
Original         & -10.35  & -4.31 & -4.37 & -10.35 & -4.72 & -4.06  \\
 Add $\boldsymbol{-p}$ to Real & 17.85 & 4.54 & 2.99 & 17.85 & -0.28  & -1.17 \\ \toprule
\end{tabular}
}
\label{add}
\end{table}

This section aims to use experimental results to support three characteristics in the causes of invisible relevance bias mentioned above. 

For the first point, human evaluation in Section~\ref{human_evaluation} and retrieval performance in Section~\ref{retrieval_performance} have shown that AI-generated images do not introduce additional visual semantics compared to their real images, indicating that the additional information is invisible. Besides, the ranking bias detected in Section~\ref{ibb} and Section~\ref{training_bias} shows that this additional information can be embedded by the image encoder and produce higher relevance score than real images. 

For the second point, we design a direct experiment to support it. We apply the bitwise addition of the reverse transformations vector $\boldsymbol{-p}$ to the representations of real images encoded by the original, non-debiased retriever and detect whether the bias can be eliminated. The intention for this is that if this additional information ($\boldsymbol{-p}$) is indeed the cause of the higher ranking of AI-generated images, then by incorporating this information into the representation of real images, the real images will similarly attain a higher ranking. Consequently, this would mitigate the ranking disparity between real and generated images. The experimental results are shown in Table~\ref{add}. It is very surprising that the ranking advantage of generated images over real images caused by invisible relevance bias is not only eliminated but reversed by simply bitwisely adding $\boldsymbol{-p}$ to the representation of the real images without any debiasing training. This proves that the reverse transformations vector $\boldsymbol{-p}$ we found is an important cause of the invisible relevance bias. It is implicit in the AI-generated images and can be embedded into the representations by the image encoder.

For the third point, T-SNE visualization of image representations and transformations vector $\boldsymbol{p}$ in Figure~\ref{tsne_casue} has shown that compared with the scattered image representations, the transformations vector $\boldsymbol{p}$ show an obvious aggregation phenomenon. This proves that for AI-generated images with different semantics, the debiased model only needs roughly consistent transformations on representations to remove the bias, which means that there is a certain consistency in the additional information for AI-generated images encoded by the image encoder.


\section{Related Work}
\subsection{Artificial Intelligence Generated Content}
With the development of generation models, AI-generated content (AIGC) becomes more realistic and less discernible~\cite{agnese2020survey,cao2023comprehensive,wu2023ai}. The use of AI to generate content has greatly changed the way of
content generation. It reduced the cost of content generation and improved the efficiency than traditional compared to traditional content generation pipeline biased on humans~\cite{hanley2023machine,spitale2023ai}. For example, Stable Diffusion~\cite{diffusion,croitoru2023diffusion} and DALL-E-3~\cite{betker2023improving} can generate high-quality images by simply following a short text description of uses' input. However, as AIGC is gradually applied to various content production pipelines in society and the internet, some potential risk concerns are also worthy of consideration. Some studies have found that AI-generated content has risks in many aspects, such as discrimination, privacy leakage, ethics, and safety~\cite{deshpande2023toxicity,zhuo2023exploring,chen2023can,jiang2023disinformation,su2023fake}. A recent study finds that texts generated by AI introduce bias in text retrieval, causing the retriever to rank them higher~\cite{dai2023llms}. This paper extends the study of this bias to cross-modal retrieval.

\subsection{Bias and Fairness in Information Retrieval}

Due to the influence of training data, indexes, models, etc., the ranking results of information retrieval will produce certain deviations~\cite{gao2021addressing}. These biases are specifically reflected in various aspects such as gender~\cite{azzopardi2021cognitive}, cognition~\cite{bigdeli2021exploring}, understandability~\cite{zuccon2016understandability}, and retrievability~\cite{wilkie2014retrievability}. Kulshrestha et al.~\cite{kulshrestha2017quantifying} formulated a framework for quantifying bias, addressing its origins in both the data source and the ranking system. Beutel et al.~\cite{beutel2019fairness}, Kuhlman et al.~\cite{kuhlman2019fare}, and Yao and Huang~\cite{yao2017new} introduced pairwise comparisons focusing on utility and prediction errors. Geyik et al.~\cite{geyik2019fairness} and Yang along with Stoyanovich~\cite{yang2017measuring} suggested fairness measures based on distances for ranked outputs, grounded in statistical parity. Gao~\cite{gao2021toward} advocated for a unified evaluation metric tailored for fairness-aware ranking algorithms. Diaz et al.~\cite{diaz2020evaluating} proposed metrics specifically designed for assessing fairness in stochastic rankings. Numerous studies have conducted comparisons among existing fairness metrics~\cite{geyik2019fairness,diaz2020evaluating,raj2020comparing,sapiezynski2019quantifying}. In this paper, we focus on the ranking bias caused by AI-generated Images in text-image retrieval.

\section{Conclusion}
This paper explores the impact on text-image retrieval introduced by AI-generated images. We construct a reasonable benchmark to simulate the retrieval scenarios comprising both real images and AI-generated images. Experiments on this benchmark underscore that AI-generated images tend to be ranked higher by retrieval models, despite lacking more visually relevant semantics to the queries than real images. We define this bias as \textbf{invisible relevance bias}. This bias is prevalent across retrieval models with varying training data and architectures. Moreover, mixing AI-generated images into the training data makes the bias more serious, causing a vicious cycle where AI-generated images gain more exposure from massive data, increasing their likelihood of being mixed into retrieval model training, and exacerbating the bias further. To solve this, we propose an effective debiasing method to mitigate the invisible relevance bias. Then, we use our proposed debiasing method to reversely determine that the cause of invisible relevance is that the AI-generated images cause the image encoder to introduce additional information into their representation. This additional information can make the retriever estimate a higher relevance score. Findings in this paper reveal the potential impact of AI-generated images on text-image retrieval systems in the context of the rapid development of AIGC and have implications for further research.

\begin{acks}
This work was supported by the National Key R\&D Program of China (2022YFB3103700, 2022YFB3103704), the National
Natural Science Foundation of China (NSFC) under Grants No. 62276248 and U21B2046, and the Youth Innovation Promotion Association
CAS under Grants No. 2023111.
\end{acks}


\normalem

\bibliographystyle{ACM-Reference-Format}
\bibliography{my_new}

\end{document}